\documentclass[aps,pre,preprint,onecolumn,citeautoscript,superscriptaddress]{revtex4-1}  
\usepackage{amsmath,amssymb,bm} 
\usepackage{graphicx}  
\usepackage[tight]{subfigure}    
\usepackage{color} 
\usepackage{mathtools}
\usepackage{multirow}
\usepackage[papersize={8.5in,11in}]{geometry}
\usepackage[colorlinks=true]{hyperref}  
\hypersetup{
    bookmarks=true,         
    unicode=false,          
    pdftoolbar=true,        
    pdfmenubar=true,        
    pdffitwindow=false,     
    pdfstartview={FitH},    
    pdftitle={My title},    
    pdfauthor={Author},     
    pdfsubject={Subject},   
    pdfcreator={Creator},   
    pdfproducer={Producer}, 
    pdfkeywords={keyword1} {key2} {key3}, 
    pdfnewwindow=true,      
    colorlinks=true,       
    linkcolor=magenta, 
    citecolor=blue,        
    filecolor=magenta,      
    urlcolor=cyan           
} 

\newcommand{\beq}{\begin{equation}}
\newcommand{\eeq}{\end{equation}}
\newcommand{\nn}{\nonumber \\}
\def\bea{\begin{eqnarray}}
\def\eea{\end{eqnarray}}

\begin{document}
\title{Transition from the $\mathbb{Z}_2$ spin liquid to antiferromagnetic order:\\ spectrum on the torus}  
\author{Seth Whitsitt}
 \affiliation{Department of Physics, Harvard University, Cambridge, Massachusetts, 02138, USA}
 \author{Subir Sachdev}
 \affiliation{Department of Physics, Harvard University, Cambridge, Massachusetts, 02138, USA}
 \affiliation{Perimeter Institute for Theoretical Physics, Waterloo, Ontario N2L 2Y5, Canada}
 \date{Nov 1, 2015\\
 \vspace{0.6in}}
\begin{abstract} 
We describe the finite-size spectrum in the vicinity of the quantum critical point between a $\mathbb{Z}_2$ spin liquid and a coplanar antiferromagnet on the torus. We obtain the universal evolution of all low-lying states in an antiferromagnet with global SU(2) spin rotation symmetry, as it moves from the 4-fold topological degeneracy in a gapped $\mathbb{Z}_2$ spin liquid to the Anderson ``tower-of-states'' in the ordered antiferromagnet.
Due to the existence of nontrivial order on either side of this transition, this critical point cannot be described in a conventional Landau-Ginzburg-Wilson framework. Instead it is described by a theory involving fractionalized degrees of freedom known as the O$(4)^\ast$ model, whose spectrum is altered in a significant way by its proximity to a topologically ordered phase. We compute the spectrum by relating it to the spectrum of the O$(4)$ Wilson-Fisher fixed point on the torus, modified with a selection rule on the states, and with nontrivial boundary conditions corresponding to topological sectors in the spin liquid. The spectrum of the critical O($2N$) model is calculated directly at $N=\infty$, which then allows a reconstruction of the full spectrum of the O($2N)^\ast$ model at leading order in $1/N$. This spectrum is a unique characteristic of the vicinity of a fractionalized quantum critical point, as well as a universal signature of the existence of proximate $\mathbb{Z}_2$ topological and antiferromagnetically-ordered phases, 
and can be compared with numerical computations
on quantum antiferromagnets on two dimensional lattices.
\end{abstract}
\maketitle 
\section{Introduction}
\label{sec:intro}

Recent numerical studies \cite{White15,Sheng15} of the spin $S=1/2$ antiferromagnet on the triangular lattice have 
presented convincing evidence for a spin liquid ground state in the presence of  a next-nearest neighbor exchange interaction ($J_2$).
They also find an apparently continuous transition to an antiferromagnetically ordered ground state at smaller $J_2$, with the familiar
3-sublattice coplanar order of the triangular lattice. Here we will assume that this antiferromagnetic state is the same as the conventional state described by the semiclassical spin-wave theory, and possesses only integer spin excitations. So the transition from the spin liquid to the antiferromagnet is a confinement transition, associated with the confinement of half-integer spin excitations.

An attractive candidate for the observed spin liquid is the $\mathbb{Z}_2$ spin liquid \cite{NRSS91,RJSS91,XGW91,SSkagome}.
The purpose of our paper is to examine a confinement transition of the $\mathbb{Z}_2$ spin liquid on a torus geometry.
The torus is characterized by a length scale, a circumference $L$, and a modular parameter $\tau$. At a continuous quantum phase transition associated with a conformal field theory, the low-lying quantum states on a torus have an energy proportional to $c/L$
(where $c$ is a spin-wave velocity, which will henceforth be set to unity), with proportionality constants which are universal functions of $\tau$. We will show that this torus spectrum contains characteristic signatures of the topological order of the proximate $\mathbb{Z}_2$ spin liquid. The spectrum exhibits a universal crossover from the 4-fold topological degeneracy of the 
$\mathbb{Z}_2$ into the characteristic spectrum of the confining phase: in our case the confining phase has long-range
antiferromagnetic order and has a low-lying Anderson ``tower-of-states'' \cite{PWA52,Bernu92,Bernu94}
signaling the spontaneous breaking of the SU(2) spin-rotation symmetry.
It is our hope that these results on the spectrum of the critical point will aid numerical studies of quantum antiferromagnets,
and help identify the topological order of proximate spin liquid phases.

A theory of a confinement transition of the $\mathbb{Z}_2$ spin liquid was initially presented in Refs.~\onlinecite{RJSS91,MVSS99}, 
in terms of a frustrated
Ising model obtained from an `odd' dimer model; the same theory appeared later in other models \cite{TSMPAF99,MS01,MSF02}, 
and in recent work \cite{HHK15,WQCM16,PCAS16}.
This confinement transition can be interpreted in terms of the condensation of the $m$ particle 
(the `vison') of the toric code \cite{AYK97}, but with the modification that the $m$ particle carries non-trivial quantum numbers of the space group of the underlying 
lattice (in modern terminology, the `odd' dimer model realizes a `symmetry enriched topological' (SET) state \cite{WenSPT}). 
The non-trivial quantum numbers of the vison lead to lattice symmetry-breaking in the confining state. 
In other models \cite{MTS02}, including the toric code and `even' dimer models, the $m$ particles transforms trivially under the space
group, and then the confining state does not break any symmetries. A theory for the finite-size spectra across such a non-symmetry-breaking
confinement transition, along with exact diagonalization results in a model system, appear in a separate paper \cite{HLSSW16}.

In the present paper, we are interested in the case where the condensing particle carries half-integer quantum numbers of the total spin, 
which is the `spinon' of the spin liquid (conventionally labeled as analogous to the $e$ particle of the toric code).
A theory for the condensation of spinons from the SET state of a $\mathbb{Z}_2$ spin liquid on the triangular lattice 
was presented in Refs.~\onlinecite{CSS93,CSS94}, 
and this theory will form the basis of our computations here. The order parameter of the 
coplanar antiferromagnet 
is identified by points on the SO(3) manifold, and so the Landau-Ginzburg-Wilson (LGW) framework suggests a field theory based on such an order parameter. However, the theory of Refs.~\onlinecite{CSS93,CSS94} is a `deconfined' critical theory 
beyond the LGW paradigm, and is instead
expressed in terms of a spinon field which is identified by points on SU(2)$\equiv S_3$.

The connection between coplanar magnetic order and the spinon in the spin liquid phase can be made explicit. We write the expectation value in the ordered state as
\beq
\langle \mathbf{S}_j \rangle = S \left[ \mathbf{n}_1 \cos\left( \vec{Q} \cdot \vec{x}_j \right) + \mathbf{n}_2 \sin \left( \vec{Q} \cdot \vec{x}_j \right) \right]
\eeq
where the ordering wave vector is $\vec{Q} = 4 \pi\left( 1/3,1/\sqrt{3} \right)$ for the semiclassical ground state of the Heisenberg model on the triangular lattice. The vectors $\mathbf{n}_{1,2}$ are arbitrary up the constraints
\beq
\mathbf{n}_1^2 = \mathbf{n}_2^2 = 1, \qquad \mathbf{n}_1 \cdot \mathbf{n}_2 = 0 \label{constr}
\eeq
Different orientations of these two vectors are related by a rotation matrix, identifying the order parameter as an element of SO(3). A conventional LGW description of a transition from this magnetically ordered state to a paramagnetic state would begin with an effective action for the fluctuations of the vectors $\mathbf{n}_{1,2}$. However, this phase transition would drive the system into a trivial gapped paramagnetic state with a non-degenerate ground state, which cannot occur in a system with an odd number of half-integer spins per unit cell such as the triangular antiferromagnet \cite{hastings04}. Therefore, we seek a description in terms of fractionalized degrees of freedom. Following Refs.~\onlinecite{CSS93,CSS94}, we write
\beq
n_{1 a} + i n_{2 a} = \sum_{\alpha, \beta, \gamma = 1}^2 \epsilon_{\alpha \gamma} z_{\gamma} \sigma^{a}_{\alpha \beta} z_{\beta} \label{spinors}
\eeq
This parametrization explicitly solves the constraints in Eq.~(\ref{constr}), and it can be checked that the complex bosonic field $z_{\alpha}$, with $\alpha = \uparrow, \downarrow$, transforms as an $S=1/2$ spinor under spin rotations. However, this representation is doubled-valued: one can perform a gauge transformation, $z_{\alpha}(\mathbf{x},\tau) \rightarrow \eta(\mathbf{x},\tau) z_{\alpha}(\mathbf{x},\tau)$, $\eta = \pm 1$, at any point in space-time and obtain an equivalent representation of the physically observable order parameter. This identifies the order parameter space as SU(2)/$\mathbb{Z}_2$, which is equivalent to SO(3). This description is complementary to the confinement transition described above, where $z_{\alpha}$ is identified with the SU(2) spinon of the $\mathbb{Z}_2$ spin liquid. We note that as the spinon condenses, the only remnant of the gapped vison in the spin liquid is the double-valued nature of the spinon field.

We therefore write a critical theory for the complex boson $z_\alpha$, taking values in SU(2), consistent with the allowed symmetries. 
Keeping only terms relevant at the critical point, the universal Lagrangian of the transition in 2+1 dimensional spacetime is 
\beq
\mathcal{L} = | \partial_\mu z_\alpha|^2 + s |z_\alpha|^2 + u \left( |z_\alpha|^2 \right)^2 . \label{L}
\eeq
The `mass' $s$ has to be tuned to a critical value $s=s_c$ to access the critical point, while $u$ approaches a non-zero value
determined by the Wilson-Fisher fixed point \cite{WF72}. Note that this spin-1/2 relativistic boson is not in contradiction with the
spin-statistics theorem, because here `spin' refers to a global flavor symmetry, rather than the intrinsic angular momentum of relativistic particles. 
We will allow the index $\alpha$ to range over $1 \ldots N$, 
and obtained results in the $1/N$ expansion. Note that the theory $\mathcal{L}$ has O($2N$) symmetry, and so we
will be examining properties of the O($2N$) fixed point.

A first guess towards obtaining the spectrum on the torus is that we simply have to solve the theory $\mathcal{L}$ on the torus
with periodic boundary conditions on the spinon field $z_\alpha$:
\beq
z_\alpha (x+iy + n_1 \omega_1 + n_2 \omega_2 )  = z_\alpha (x+i y) \label{pbc}
\eeq
where $x,y$ are the spatial co-ordinates, $n_{1,2}$ are integers, and $\omega_{1,2}$ are the complex periods of the torus with $\tau = \omega_2/\omega_1$; we choose $|\omega_1| = L$. 
These boundary conditions would be appropriate if we were solving for the spectrum
of an O($2N$) rotor model, for a transition from an ordered state with $\left\langle z_\alpha \right\rangle \neq 0$ to a trivial
paramagnet with $\left\langle z_\alpha \right\rangle = 0$. 

However, in our case we are considering a transition to a paramagnet with
$\mathbb{Z}_2$ topological order, and this does have important consequences for the spectrum of the critical theory. 
A first consequence follows from the fact that no physical operator can be associated with a single $z_\alpha$ operator, and all observables involve at least bilinears of $z_\alpha$ and $z_\alpha^\ast$. The periodic boundary conditions on the torus apply to the physical spin operators of the antiferromagnet, and so for the spinons we have the more general boundary conditions \cite{HLSSW16}
\beq
z_\alpha (x+iy + n_1 \omega_1 + n_2 \omega_2 )  = \pm z_\alpha (x+i y) . \label{sbc}
\eeq
The anti-periodic boundary conditions correspond to the presence of a vison flux in the corresponding cycle of the torus.
In the $\mathbb{Z}_2$ spin liquid, such boundary conditions lead to the near four-fold degeneracy of the ground state, with the states differing by an energy which is exponentially small in $L$. At the quantum critical point, this degeneracy evolves into additional
states which are spaced by an energy of order $1/L$. 

A second consequence arises from the fact that the all states share the same number of spinons modulo 2. In other words, 
if the underlying lattice antiferromagnet has an even (odd) number of $S=1/2$ spins on the torus, then all states will carry
integer (half-integer) spin. This implies that the wavefunctional, $\Psi$, of the critical theory obeys \cite{HLSSW16}
\beq
\Psi [ - z_\alpha (x+iy) ] = \, \eta \, \Psi [z_\alpha (x + i y) ] \label{evenodd}
\eeq
where $\eta = +1$ $(-1)$ for an even (odd) number of spins. We postulate here that there is a universal spectrum
at the critical point between the $\mathbb{Z}_2$ spin liquid and an ordered antiferromagnet which is described by the 
O($2N$) critical theory in Eq.~(\ref{L}) subject to the boundary condition in Eq.~(\ref{sbc}) and the 
constraint in Eq.~(\ref{evenodd}).
Following the notation of Ref.~\onlinecite{TSMPAF99}, we will call this the O$(2N)^\ast$ critical theory, while the theory
obeying the boundary condition in Eq.~(\ref{pbc}) is the conventional O$(2N)$ theory. It was previously pointed out \cite{BSTS12} 
that the
O$(2N)^\ast$ critical theory has a distinct entanglement entropy from the O$(2N)$ theory; our results show that the distinction
also applies to the finite-size spectrum on a torus. 

In the application to the lattice antiferromagnet, we also have to consider the fact that the O$(2N)$ symmetry of $\mathcal{L}$
is an emergent symmetry of the critical point, and is not a symmetry of the underlying Hamiltonian. So we have to consider
operators which break the O$(2N)$ symmetry. All operators which break the O($2N$) symmetry down to SU($N$) are 
irrelevant at the critical point, and we will consider here only the leading irrelevant operator. This is given by \cite{CSS93,CSS94}
\beq
\mathcal{L}' = \gamma |z_\alpha^\ast \nabla z_\alpha|^2 \label{Lp}
\eeq
We also describe the leading perturbative effect of $\gamma$ on the critical spectrum.

We will begin in Section \ref{sec:onspec} by a description of the torus spectrum of the O($2N$) critical theory in the $1/N$ expansion, followed by a discussion of the evolution of the spectrum between the ordered and disordered phases in Section \ref{secevo}.
This will be followed by the corresponding results for the O$(2N)^\ast$ critical theory in Section \ref{ostar} as well as a discussion of new features of the spectrum in the topological and ordered phases in Section \ref{evotopo}. 
The effects of $\mathcal{L}'$ will be considered in Section \ref{aniso}.

\section{Critical O($2N$) spectrum: large N}
\label{sec:onspec}
\subsection{General formalism}
\label{largen}
In this section, we develop our formalism for the large-$N$ expansion of the critical O($2N$) model. For a review of the large-$N$ expansion, see Ref.~\onlinecite{poly87}. We take the Euclidean action
\beq
\mathcal{S} = \int d\tau d^2 x \left(| \partial_\mu z_\alpha|^2 + u s |z_\alpha|^2 + \frac{u}{2 N} \left( |z_\alpha|^2 \right)^2 \right) . \label{S}
\eeq
We choose a slightly different notation for the couplings compared with Eq.~(\ref{L}), which will simplify subsequent expressions. We will perform the large $N$ expansion at fixed $u$, and tune the quadratic coupling to its critical value $s = s_c$. Subsequently we will take the $u \rightarrow \infty$ limit in each term
to obtain the scaling limit. We will also consider deviations from the critical coupling $s - s_c$.

\begin{figure}
\centering
\includegraphics[width=11cm]{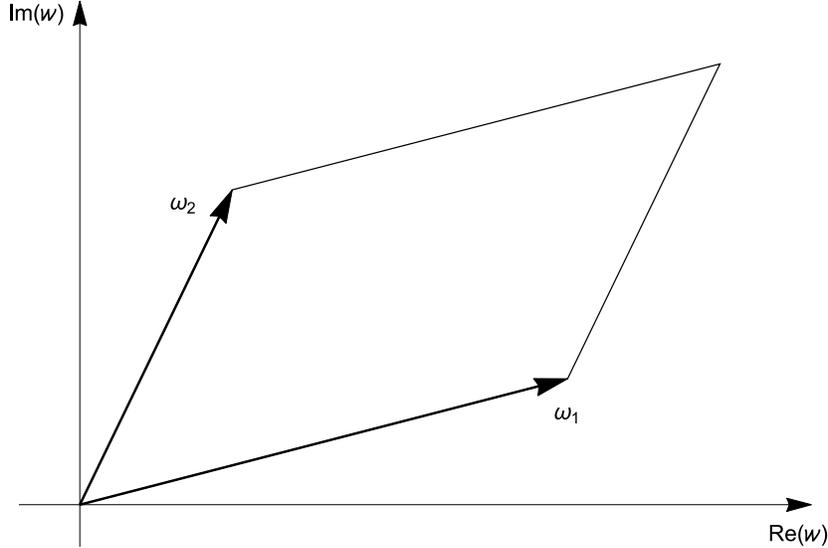}
\caption{The geometry of the spatial torus, where the position is given by complex coordinates $w = x + iy$. We associate all points related by a lattice vector $n\omega_1 + m \omega_2$ for $n,m \in \mathbb{Z}$, where the complex numbers $\omega_1$ and $\omega_2$ are the periods of the torus. We define the modular parameter $\tau = \omega_2/\omega_1$ and the length scale $L \equiv |\omega_1| = \sqrt{\mathcal{A}/\mathrm{Im}(\tau)}$.}
\label{torusfig}
\end{figure}

The field theory is defined on a spatial torus, which can be parametrized by complex coordinates $w = x + iy$. The torus is defined by two complex periods $\omega_1$ and $\omega_2$, an area $\mathcal{A} = \mathrm{Im}(\omega_2\omega_1^\ast)$, and we define the dimensionless modular parameter $\tau = \omega_2/\omega_1$ with real and imaginary parts denoted $\tau = \tau_1 + i \tau_2$. The geometry is shown in Fig. \ref{torusfig}. We also define the length scale $L \equiv |\omega_1| = \sqrt{\mathcal{A}/\tau_2}$. In this geometry, the basis vectors of the dual lattice are given by
\beq
k_1 = -i\omega_2/\mathcal{A}, \qquad k_2 = i \omega_1/\mathcal{A},
\label{dualk}
\eeq
so a general momentum vector takes the form
\beq
k_{n,m} =  2 \pi \left(n k_1 + mk_2\right), \qquad n, m \in \mathbb{Z}.
\eeq

We can rewrite the path integral (up to an unimportant constant) as
\beq
Z = \int \mathcal{D} z_{\alpha} \exp \left( -\int d^2x d \tau \left[ | \partial_\mu z_\alpha|^2 + \frac{u}{2 N} \left( |z_\alpha|^2 + \frac{N s}{2} \right)^2 \right] \right). \label{S2}
\eeq
We decouple the quartic term by introducing an auxiliary field $\tilde{\lambda}$:
\beq
Z = \int \mathcal{D} z_{\alpha} \mathcal{D}\tilde{\lambda} \exp\left( -\int d^2x d \tau \left[ | \partial_\mu z_\alpha|^2 + i \tilde{\lambda} \left( |z_\alpha|^2 + \frac{N s}{2} \right) + \frac{N \tilde{\lambda}^2}{2 u} \right] \right). \label{S3}
\eeq
The $z_\alpha$ can be integrated out, obtaining an action for $\tilde{\lambda}$,
\beq
Z = \int \mathcal{D} \tilde{\lambda} \exp \left[ - N \ \mathrm{Tr} \ln\left( - \partial_{\tau}^2 - \nabla^2 + i \tilde{\lambda} \right) - N \ \int d\tau d^2x \left( \frac{\tilde{\lambda}^2}{2 u} + \frac{s}{2} i \tilde{\lambda} \right)\right].
\eeq
At $N = \infty$, we should expand around the saddle point value, which we call $i \tilde{\lambda} = \Delta^2$, and is given by
\beq
\frac{\Delta^2}{u} = \frac{s}{2} + \frac{1}{\mathcal{A}} \sum_k \int \frac{d \omega}{2 \pi} \frac{1}{\omega^2 + |k|^2 + \Delta^2}. \label{gap1}
\eeq

At this point we tune $s \rightarrow s_c$ such that the correlation length diverges when $\mathcal{A} \rightarrow \infty$. From Eq.~(\ref{S3}), it is clear that the correlation length at $N = \infty$ is just the inverse of $\Delta$, so $s_c$ is
\beq
s_c = - 2 \int \frac{d \omega}{2 \pi} \frac{d^2 k}{4 \pi^2} \frac{1}{(\omega^2 + |k|^2)} = - 2 \int \frac{d^2 k}{4 \pi^2} \frac{1}{2 k}.
\eeq
We can add and subtract $s_c$ from Eq.~(\ref{gap1}) while taking the limit $u \rightarrow \infty$, and we find
\beq
s - s_c = \int \frac{d^2 k}{4 \pi^2} \frac{1}{k} - \frac{1}{\mathcal{A}} \sum_{k} \frac{1}{\sqrt{|k|^2 + \Delta^2}}. \label{D0}
\eeq
This equation is to be solved for $\Delta$, yielding an answer of the form $\Delta = \#/L$, where $\#$ is a universal function of $L(s - s_c)$ independent of the regularization scheme at large momenta. From the general theory of finite-size scaling \cite{BZJ85}, the energy levels should take the form 
\beq
E_n = \frac{1}{L}X_n\left[L^{1/\nu}(s - s_c)\right],
\eeq 
for some universal set of functions $X_n$, so our expressions show that $\nu = 1$ at $N=\infty$ in $(2+1)$-dimensions. 

In this paper, we use dimensional regularization to evaluate divergent sums, which sets $s_c = 0$. The computation is given in Appendix \ref{dimreg}, and in terms of the special functions defined there, the gap equation becomes 
\beq
g^{(2)}_{1/2}(\Delta,\tau) = - 2 \pi L (s - s_c), \label{gapeqn}
\eeq
which is solved numerically. At the critical point, $s = s_c$, the gap $\Delta$ depends only on the geometry of the torus. We note that $\Delta$ is a monotonically increasing function of $(s - s_c)$.

We also find the ground-state energy. This is computed from the path integral by temporarily taking a finite length in the time-direction, $0 < t < T$, and then taking the limit 
\beq
E_0 = - \lim_{T \rightarrow \infty} \frac{1}{T}\ln Z.
\eeq
Directly taking $i \tilde{\lambda} = \Delta^2$ and $u = \infty$, this is given by
\bea
E_0 &=& N \sum_k \int \frac{d \omega}{2 \pi} \ln \left( \omega^2 + |k|^2 + \Delta^2 \right) + \frac{N s }{2} \mathcal{A} \Delta^2 \nn 
&=& N \sum_k \int \frac{d \omega}{2 \pi} \ln \left( \omega^2 \right) + N \sum_k \sqrt{|k|^2 + \Delta^2} + \frac{N s }{2} \mathcal{A} \Delta^2.
\label{egs}
\eea
We subtract the first term, which is independent of the system size and boundary conditions. The remaining sum is evaluated using dimensional regularization,
\beq
E_0 = \frac{2 \pi N}{\tau_2 L} g_{-1/2}^{(2)}(\Delta,\tau) + \frac{N (s - s_c) }{2} \tau_2 L^2 \Delta^2, \label{egs2}
\eeq 
where the special function $g_{-1/2}^{(2)}(\Delta,\tau)$ is defined in Eq.~(\ref{specfunc}). Our choice of renormalization has set $E_0 = 0$ at $s = s_c$ and $L = \infty$, where the theory has full conformal invariance.

Now that we have the saddle point value of $\tilde{\lambda}$ at $N = \infty$, we can read off the Euclidean-time propagator of $z_{\alpha}$
\beq
G_0(k,i\omega) \equiv \int d^2x d\tau e^{-i x k - i \omega \tau} \langle z_\alpha(x,\tau) z^\dagger_\beta(0,0) \rangle = \frac{\delta_{\alpha \beta}}{\omega^2 + k^2 + \Delta^2}.
\eeq
We also expand in the fluctuations of $\tilde{\lambda}$. Writing $i \tilde{\lambda} = \Delta^2 + i \lambda/\sqrt{N}$, the effective action is
\bea
Z &=& \int \mathcal{D}\lambda \exp\left( -\mathcal{S}_0 - \mathcal{S}_1 \right), \nn
\mathcal{S}_0 &=& \frac{1}{2\mathcal{A}} \sum_k \int \frac{d \omega}{2 \pi} \left( \Pi(k,\omega) + \frac{1}{u} \right) \lambda^2  \label{efflam}
\eea
with
\bea
\Pi (k, i \omega) &=& \frac{1}{\mathcal{A}} 
\sum_{q} \int \frac{d \Omega}{2 \pi} \frac{1}{(\Omega^2 + |q|^2 + \Delta^2) ((\omega + \Omega)^2 + |k+q|^2 + \Delta^2)} \nn
&=& \frac{1}{\mathcal{A}} 
\sum_{q} \frac{\sqrt{|q|^2 + \Delta^2} + \sqrt{|k+q|^2 + \Delta^2}}{2 \sqrt{(|q|^2 + \Delta^2)(|k+q|^2+ \Delta^2)} ( (\sqrt{|q|^2 + \Delta^2} + \sqrt{|k+q|^2 + \Delta^2})^2 + \omega^2)} \label{Pi}
\eea
and $\mathcal{S}_1$ contains nonlinear terms. We discuss $\mathcal{S}_1$ and $1/N$ corrections in Appendix \ref{1/n}. We see that the $\lambda$ propagator at $N=\infty$ is
\beq
D_0(k,i\omega) \equiv \int d^2x d\tau e^{i x k + i \omega \tau} \langle \lambda_\alpha(x,\tau) \lambda^\dagger_\beta(0,0) \rangle = \frac{1}{\Pi(k,i\omega) + 1/u}.
\label{prop}
\eeq
This is related to the propagator of $|z_\alpha|^2$. This is most easily seen directly from the action (\ref{S3}), where $\lambda$ is not a dynamical field. Integrating out the field $i \lambda$ is equivalent to replacing it by its equation of motion,
\beq
i \lambda = \frac{u}{\sqrt{N}} | z_{\alpha}|^2 + \sqrt{N} \left( \frac{u s}{2} - \Delta^2 \right).
\eeq
So the propagator of $\lambda$ is related to the propagator of $|z_{\alpha}|^2$ by
\beq
\langle |z_{\alpha}|^2(x,\tau) |z_{\alpha}|^2(0) \rangle_c = -\frac{N}{u^2} \langle \lambda(x,\tau) \lambda(0) \rangle.
\eeq
This can also be verified by coupling a source $J$ to $|z_{\alpha}|^2$ and taking functional derivatives \cite{DPSS12}.

\subsection{Spectrum}
\label{spec}
We describe the spectrum in terms of ``$n$-particle states,'' which are created by $n$ fields:
\beq
\underbrace{b^{\dagger}_{\alpha} b^{\dagger}_{\beta} \cdots b^{\dagger}_{\gamma}}_{n} | 0\rangle. \label{particles}
\eeq
The enlarged O($2N$) symmetry rotates spinons into anti-spinons, so we define $b_{\alpha}^{\dagger}$ with indices running from $\alpha = 1,..., 2N$ which can create either particle. The single-particle states are created by a single $z$ field, so by the form of the $z$ propagator, their energy is given by the Hamiltonian
\begin{equation}
H_0 = E_0 + \sum_{k \alpha} \sqrt{|k|^2 + \Delta^2} b^{\dagger}_{\alpha}(k)b_{\alpha}(k),
\label{eqn:ham}
\end{equation}
where $\alpha = 1, ..., 2N$. The energy of the state $b_{\alpha}^{\dagger}(k)|0\rangle$ is given by 
\beq
E_1(k) = E_0 + \sqrt{|k|^2 + \Delta^2}. \label{onep}
\eeq 
This state is in the fundamental representation of O($2N$), so it is $2N$-fold degenerate in addition to any degeneracies between values of $k$.

Two particle states with momentum $k$ take the form 
\beq
b^{\dagger}_{\alpha}(q) b_{\beta}^{\dagger}(k-q)|0\rangle
\eeq
for all choices of momentum $q$. We decompose this into irreducible representations of O($2N$), which must separately have definite energy:
\beq
b^{\dagger}_{\alpha} b^{\dagger}_{\beta} = \delta_{\alpha \beta}\left(\frac{1}{2N} b^{\dagger}_{\gamma} b^{\dagger}_{\gamma}\right) + \bigg(b^{\dagger}_{[\alpha} b^{\dagger}_{\beta]}\bigg) + \left(b^{\dagger}_{(\alpha} b^{\dagger}_{\beta)} - \frac{\delta_{\alpha \beta}}{2N} b^{\dagger}_{\gamma} b^{\dagger}_{\gamma}\right) \equiv \delta_{\alpha \beta} S + A_{\alpha \beta} + T_{\alpha \beta}. \label{irreps}
\eeq
These are the singlet, antisymmetric tensor, and symmetric traceless tensor representations respectively. Simple counting shows that $S$ creates one state, $A_{\alpha \beta}$ creates $N(2N-1)$ states, and $T_{\alpha \beta}$ creates $(2N-1)(2N+2)/2$ states. Note that if $q = k - q$, the antisymmetric representation will not be present.

At this point we can use the analysis above. At $N = \infty$, the $z$ propagator takes the form of a free boson with dispersion $\sqrt{|k|^2 + \Delta^2}$, so one would na\"{\i}vely expect all states to have energy given by the Hamiltonian (\ref{eqn:ham}). However, this is not the case for the singlet state, since
\beq
\langle |z_{\alpha}|^2(x,\tau) |z_{\beta}|^2(0,0) \rangle \propto \langle \lambda(x,\tau) \lambda(0,0) \rangle.
\eeq
So the fact that that the propagator of $\lambda$ takes a nontrivial form at $N = \infty$ has the effect of shifting the energy of singlet states. The energies of the singlet states are given by the poles in $D(k,i\omega)$, or equivalently the zeros of $\Pi(k,i\omega)$. From Eq.~(\ref{Pi}) we see that $\Pi$ is always convergent in $d=2$, so we can sum the series numerically to find the singlet energies, which are given by
\beq
\Pi(k,E_2^{(S)}(k)) = 0. \label{singe}
\eeq
In contrast the antisymmetric tensor and symmetric traceless tensor remain degenerate at $N=\infty$, giving $4N^2 - 1$ degenerate states with energy
\beq
E_2 (k) = E_1 (q) + E_1 (k-q)
\eeq
for all choices of the momentum $q$, where $E_1 (q) $ is the single particle energy, Eq.~(\ref{onep}). The choice of $q$ can also induce additional degeneracies for any given total momentum $k$. In addition, we saw that if $q = k - q$ there will be no antisymmetric part, so there will only be a degeneracy of $(2N-1)(2N+2)/2$ from O($2N$) symmetry.

Going beyond the two-particle states, we expect that a general state will be given by an application of
\begin{equation}
b^{\dagger}_{\alpha}(k_1)b^{\dagger}_{\beta}(k_2)b^{\dagger}_{\gamma}(k_3)b^{\dagger}_{\sigma}(k_4) \cdots |0 \rangle.
\end{equation}
Past the two-particle states, the decomposition into irreducible representations becomes more involved. Generally, the states will decompose into singlets with energies given by the zeros of $\Pi(k,E(k))$, and states described by O($2N$) traceless tensors with energies given by by Fock spectrum of Eq.~(\ref{eqn:ham}). Extra degeneracies can occur due to discrete point group symmetries of the torus, and sometimes degeneracies are reduced if some of the $b^{\dagger}$s are indistinguishable. 
\begin{table}
\begin{center}
\bgroup
\def\arraystretch{1.05}
\begin{tabular}{| c || c | c | c |}

\hline
degeneracy & $\kappa = 0$ & $\kappa = 1$ & $\kappa = \sqrt{2}$ \\
\hline
1 & 0 & & \\
$2N$ & 1.512 & & \\
$(2N+2)(2N-1)/2$ & 3.024 & & \\
 \multirow{2}{*}{\resizebox{!}{16pt}{$\displaystyle \binom{2N+2}{3} - 2N$}}& \multirow{2}{*}{4.536} & & \\
  & & & \\
\multirow{2}{*}{\resizebox{!}{16pt}{$\displaystyle 2\binom{1+2N}{2N-2} - \binom{3+2N}{4}$}} & \multirow{2}{*}{6.048} & & \\
 & & & \\
$8N$ & & 6.463 & \\
\multirow{2}{*}{\resizebox{!}{16pt}{$\displaystyle 2\binom{2+2N}{2N-2} - \binom{4+2N}{5}$}} & \multirow{2}{*}{7.560} & & \\
 & & & \\
$4(4N^2 - 1)$ & & 7.975 & \\
1 & 8.126 & & \\
$8N$ & & & 9.013 \\
\multirow{2}{*}{\resizebox{!}{16pt}{$\displaystyle \ \ 2\binom{3+2N}{2N-2} - \binom{5+2N}{6} \ \ $}} & \multirow{2}{*}{9.072} & & \\
 & & & \\
 \hline
\end{tabular}
\egroup
\caption{Lowest energy splittingss $L (E - E_0)$ and their degeneracy at $s = s_c$ for large-$N$ on the square torus. The ground state energy is given by $E_0 = -.329N$. Here, $\kappa = L |k|/2\pi$.}
\label{sqspec}
\end{center}
\end{table}

\subsection{Evolution of the spectrum of a function of $s - s_c$}
\label{secevo}

In this section, we discuss the general structure of the finite-size spectrum as a function of $s - s_c$, which can be worked out on general principles in the limits $s = s_c$, $s \gg s_c$, and $s \ll s_c$. We show that our model takes the correct form in these limits before giving explicit results on the evolution of the as $s - s_c$ is varied.

\subsubsection{Critical point}

At criticality, $s = s_c$, the system at an infinite volume has full conformal invariance, and there is no scale in the theory. The excitation spectrum forms a gapless continuum, $E = k$. As a result, when the system is placed on a torus, the only possible dependence that the energy can have on the size of the system is $1/L$. Therefore, the quantities $LE$ will be universal functions of $\tau$ only. This dependence is automatic from our finite-size calculations, where the solution to the gap equation will give a pure number for $L\Delta$, and all energies manifestly have $1/L$ dependence.

\subsubsection{Disordered phase}

In the disordered phase, $s > s_c$, the system develops a gap $m$ even at $L = \infty$, and the low-energy excitations will take the form $E = \sqrt{|k|^2 + m^2}$. In the scaling limit, $m$ is of order $(s - s_c)^{\nu}$ and $\nu = 1$ at $N = \infty$. This energy gap implies that all correlations decay exponentially over a length scale $1/m \sim 1/(s - s_c)$, resulting in a very weak dependence on finite-size effects when the system is placed on a torus of size $L$, provided $L m \sim L(s - s_c) \gg 1$. Therefore, we expect the finite-size spectrum of the disordered phase to evolve to the form $E = \sqrt{|k|^2 + \Delta^2}$ at increasing $(s - s_c)$, where $\Delta = m + \mathcal{O}(e^{-Lm})$ takes the same value as it does in an infinite volume up to exponentially small corrections in $L(s - s_c)$, and the momenta $k$ are quantized according to the required boundary conditions. We also note that the threshold for singlet excitations in an infinite volume is $2m$, so the absence of large finite-size corrections suggests that the two-particle singlet spectrum will merge with the other two-particle states.

The properties of the disordered phase can be verified explicitly. By taking the $L \rightarrow \infty$ limit of Eq.~(\ref{gapeqn}), we find the exact gap in an infinite volume,
\beq
m = 2 \pi (s - s_c).
\eeq
This can be compared with the gap in a finite volume when $s \gg s_c$. In this limit, $L\Delta$ is large and we can expand $g_{1/2}^{(2)}(\Delta,\tau)$, obtaining
\beq
\Delta = 2 \pi (s - s_c) + \mathcal{O}\left(\frac{1}{L^2 (s - s_c)^2} e^{-L^2 (s - s_c)^2}\right), \qquad s \gg s_c
\eeq 
The energies of the two-particle singlet states can be verified to merge with the other two-particle states in this limit.

\begin{figure}
\centering
\includegraphics[width=17cm]{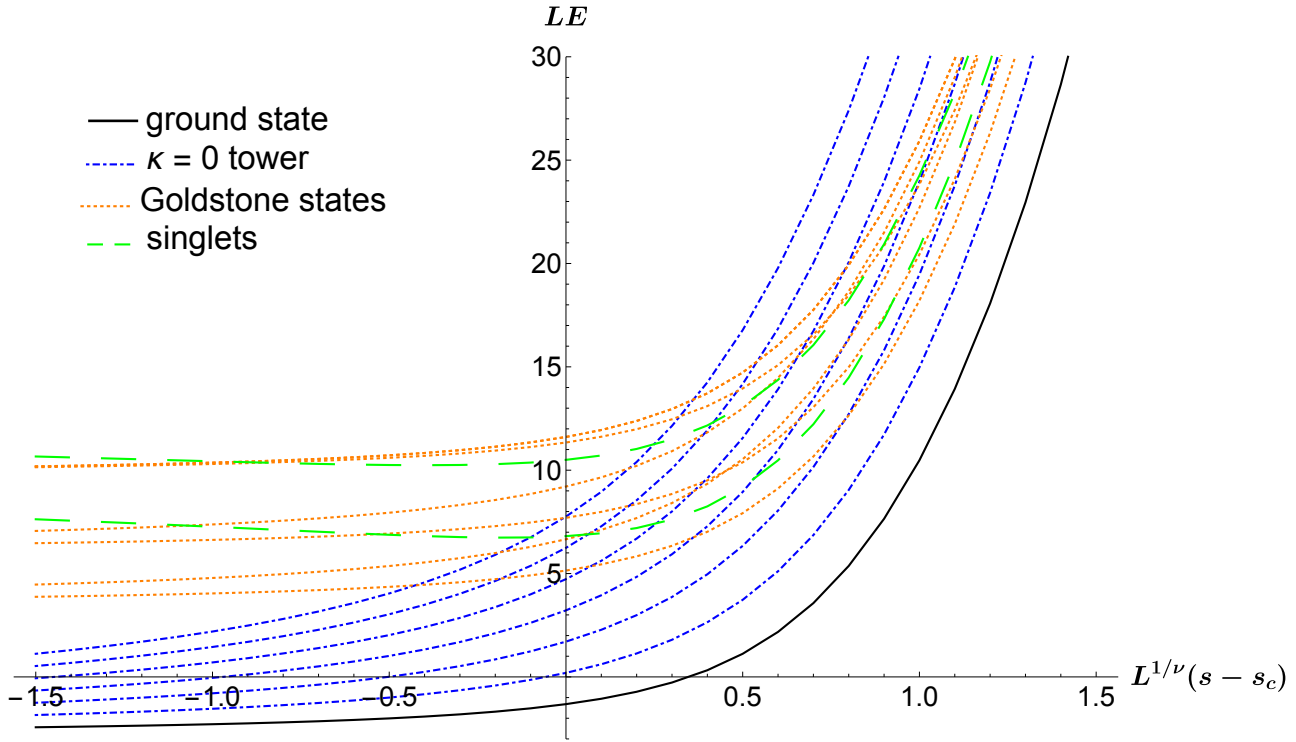}
\caption{(Color online) The evolution of the spectrum $LE$ for the O(4) model as a function of the tuning parameter $L^{1/\nu}(s - s_c)$ on the square torus, $\tau = i$. Note that $\nu = 1$ at leading order in $1/N$. The energy levels are defined so that $E = 0$ at $s = s_c$ and $L = \infty$. We label the states by their behavior in the ordered region, distinguishing between the tower, the Goldstone modes, and the singlet states. Our choice of states is not not exhaustive, but they highlight the main features of each region.}
\label{o4plot}
\end{figure}
\subsubsection{Ordered phase}

In the ordered phase, $s < s_c$, the finite-size spectrum differs considerably from the infinite volume case. In an infinite volume, there is a degenerate ground-state manifold of states at zero momentum which are related by the O($2N$) symmetry, and a properly prepared system will pick a single one of these states, spontaneously breaking the symmetry. The stable excitations above the ground state consist of $2N - 1$ Goldstone modes with a linear dispersion, $E = c |k|$, corresponding to transverse fluctuations of the order parameter about its ground state value. In addition, there will be an unstable continuum of excitations associated with transverse fluctuations of the order parameter and fluctuations of its amplitude $\phi_{\alpha}^2$, which will be mixed by interactions \cite{DPSS12}.

In contrast, in a finite volume the ground state must be a non-degenerate O($2N$) singlet, and spontaneous symmetry breaking is impossible. Instead of a ground state manifold, there will be a ``tower of states'' above the ground state at $k=0$ with energies scaling as $E \sim 1/\mathcal{A}$ with the system size \cite{PWA52,Siggia89,Bernu92,Bernu94,Azaria93,White07}. In the thermodynamic limit, this tower ``collapses'' into the ground state, and a symmetry-broken state can be formed as an extensive superposition of states in the tower. 

\begin{figure}
\centering
\includegraphics[width=8.75cm]{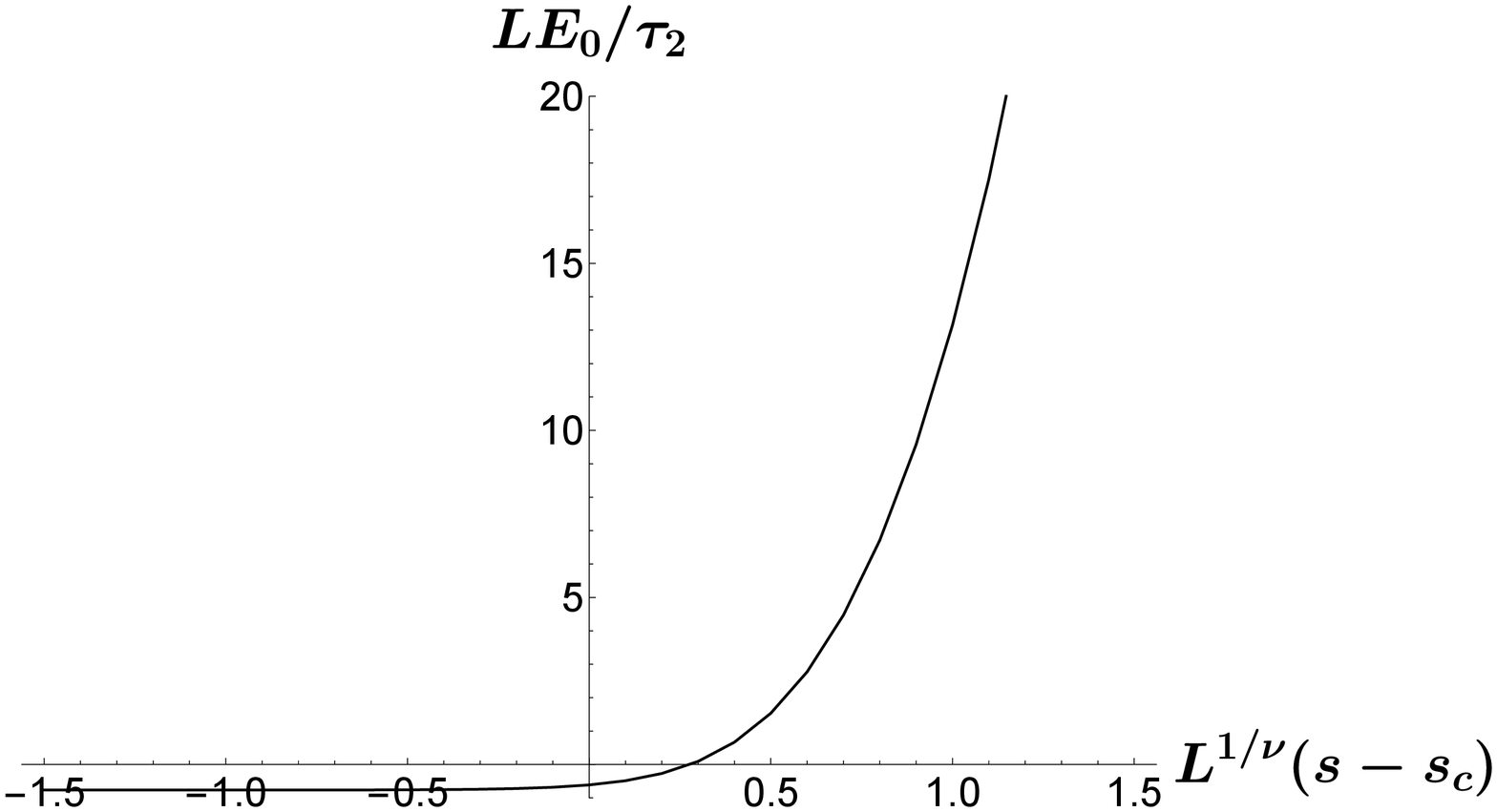}
\includegraphics[width=8.7cm]{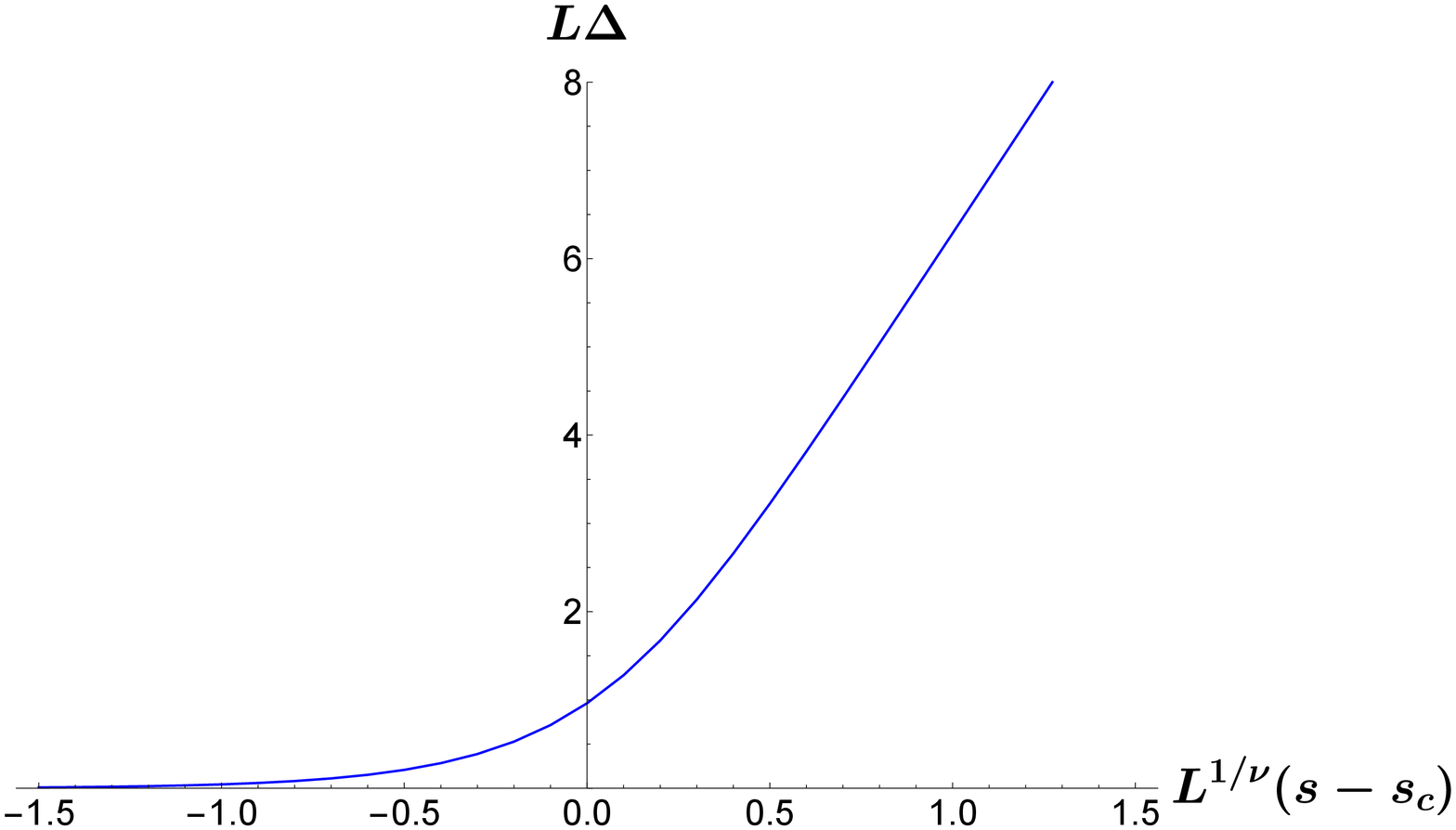}
\caption{Left: The dimensionless ground state energy density $L E_0 /\tau_2 = L^3 E_0/\mathcal{A}$ for the O(4) model on an infinite cylinder with circumference $L$. This energy is defined so that $E/\mathcal{A} = 0$ at $s = s_c$ and $L = \infty$. Right: The energy gap above the ground state for the O($2N$) model at $N = \infty$ on the infinite cylinder as a function of $s - s_c$. For energies higher than the gap, the spectrum is continuous.}
\label{cylindplot}
\end{figure}
One can analyze the general properties of the tower of states by forming an effective Hamiltonian for their spectrum. This can be derived by integrating out the finite-momentum modes and finding an effective Hamiltonian for the zero-momentum component of the field \cite{BZJ85}. For a system with O($2N)$ symmetry, the effective Hamiltonian for the tower takes the form
\beq
H_{tower} = E_0 + \frac{\mathbf{L}^2}{ \kappa\mathcal{A}N \left(s_c - s \right)} \label{towerham}
\eeq
up to corrections induced by fluctuations of the finite momentum modes. Here, $\mathbf{L}_i$, $i = 1, 2, .., N(2N-1)$ are the generators of rotations in O($2N$), and $\kappa$ is a constant which will be non-universal away from the scaling limit. The effective Hamiltonian for the tower is simply an O($2N$) rigid rotator, and the energy levels are given by
\beq
E_{\mathrm{tower}} = E_0 + \frac{ \ell (\ell + 2N - 2)}{\kappa N (s_c - s) \mathcal{A}}, \qquad \ell = 0, 1, 2, ...
\eeq
This constrains the level spacing between states in the tower. In our present calculation, we take the $N = \infty$ limit, and obtain equally-spaced energy levels. We note that for the physical cases of interest the splitting will be different; below we consider an O($4)^\ast$ transition where one takes $N = 2$ and $\ell$ even, resulting in a splitting of $2 \ell (2 \ell + 2)$ up to the irrelevant splittings discussed in Section \ref{aniso}. The eigenfunctions of Eq.~(\ref{towerham}) in the angular basis are the hyperspherical harmonics on $S^{2N-1}$, which are the higher-dimensional generalization of the familiar spherical harmonics on the two-sphere. These eigenfunctions are in the symmetric traceless tensor representations of O($2N$), and their degeneracy is given by
\beq
\mathrm{Deg.} = 2 \binom{\ell + 2N - 3}{2N - 2} + \binom{\ell + 2N - 3}{\ell}.
\eeq

We can verify the above structure in our model by taking the limit $s \ll s_c$ in the gap equation (\ref{gapeqn}). We find that the gap takes the form
\beq
\Delta = \frac{1}{\mathcal{A} (s_c - s)} + \mathcal{O}\left((\mathcal{A}(s_c - s))^{-2} \right), \qquad s \ll s_c. \label{gaptower}
\eeq
The states created purely by $|k| = 0$ will form an equally spaced spectrum above the ground state with this $1/\mathcal{A}$ dependence on the system size, and by the analysis in Section \ref{spec} they will be in the symmetric traceless tensor representations of O($2N$), in agreement with the above analysis. 

The states created by finite-momentum operators will have an energy given by $E = |k| + \mathcal{O}(\Delta^2/|k|)$, and transform in either traceless tensor or singlet representations. These correspond to the Goldstone modes in the infinite-volume system, but there will be no distinction between the longitudinal and transverse fluctuations since symmetry is unbroken. We note that even the zero-momentum states created by the singlet operator approach the expected spectrum for multi-particle Goldstone states. 

\subsection{Results}
For an explicit example, we consider the square torus, $\tau = i$, where both spatial directions have length $L$. Precisely at $s = s_c$, the energy levels are a set of universal numbers times $1/L$; in Table \ref{sqspec} we have given the lowest-lying energy levels at the critical point and their total degeneracy. We show the evolution of the spectrum $LE$ as a function of $L(s - s_c)$ in Figure \ref{o4plot}, choosing states which highlight important features of the spectrum. 

We also give results for the cylinder, $\tau_2 \rightarrow \infty$, in Figure \ref{cylindplot}. The presence of an infinite dimension changes the nature of the spectrum considerably, but there are still universal quantities to compute. Since the ground state energy is extensive, diverging with the area of the system, we plot the universal ground state energy density $L E_0/\tau_2 = L^3 E_0 / \mathcal{A}$ instead. Also, since particles can take a continuous momentum along the direction of the cylinder, the spectrum above the gap is a continuum given by particles with energy $\sqrt{k^2 + \Delta^2}$. However, the gap remains a universal quantity which we plot in Fig. \ref{cylindplot}. We also note that in the ordered phase, the gap no longer scales with $1/\mathcal{A}$ since the area is infinite. Instead, the gap becomes exponentially suppressed in the circumference of the cylinder,
\beq
\Delta \propto \frac{1}{L}\exp\left( - \pi L (s_c - s) \right), \qquad s \ll s_c, \quad \tau_2 = \infty.
\eeq

\section{Critical O$(2N)^\ast$ spectrum}
\label{ostar}
We now consider the O$(2N)^\ast$ model, where the spinons can take anti-periodic boundary conditions along either direction of the torus. We treat the four topological sectors as separate decoupled theories for now. The boundary conditions can be taken into account by simply by noticing that momentum quantization is shifted by a half-integer in the anti-periodic direction. We parametrize the momentum as
\beq
k_{n,m} =  2 \pi \left[(n + a_1) k_1 + (m + a_2)k_2\right], \qquad n, m \in \mathbb{Z}, \label{aperk}
\eeq
where the $k_i$ were defined in Eq.~(\ref{dualk}), and the values of $a_1$, $a_2$ are determined by the boundary conditions, see Table \ref{adef}.

\begin{table}
\begin{center}
\begin{tabular}{c  c}
$(\omega_1,\omega_2)$ & $(a_1,a_2)$ \\
\hline
(P,P) & $(0,0)$ \\
(P,A) & $\left(0,\frac{1}{2}\right)$ \\
(A,P) & $\left(\frac{1}{2},0\right)$ \\
(A,A) & $\left(\frac{1}{2},\frac{1}{2}\right)$ \\
\end{tabular}
\caption{The definitions of $a_1$ and $a_2$ appearing in (\ref{aperk}) for different boundary conditions. The left column denotes whether the boundary conditions are periodic (P) or anti-periodic (A) in the $\omega_1$ or $\omega_2$ directions respectively, while the right column gives the values of $a_1$ and $a_2$ for this boundary conditions.}
\label{adef}
\end{center}
\end{table}
This redefinition of allowed momenta is all that is needed to reproduce the calculations in \ref{largen}. We can still use the special functions defined in the appendix (which are defined for arbitrary boundary conditions), and we solve the same gap equation for $\Delta$,
\beq
g^{(2)}_{1/2}(\Delta,\tau) = - 2 \pi L (s - s_c), \label{agap}
\eeq
and have the same formula for the ground state energy,
\beq
E_0 = \frac{2 \pi N}{\tau_2 L} g_{-1/2}^{(2)}(\Delta,\tau) + \frac{N (s - s_c ) }{2} \tau_2 L^2 \Delta^2. \label{aenergy}
\eeq 
However, we can now find the gap and the ground state energies in all four topological sectors of the theory, and we will see below that the splitting between the ground-state energies is important. The ground-state energies are proportional to $N$, so the energy splittings in the O$(2N)^\ast$ theory will be $N$-dependent in the $1/N$ expansion, unlike the O($2N$) case above. This $N$-dependence is a physical property of a system with $2N$ spinons, since the ground state configuration of each field with a twist will each contribute equally to shift the energy above the ground state of the system without a twist.

One consequence of the anti-periodic sectors is that there is no zero mode, so the massless free particle spectrum $|k|$ already has a gap. As a result, the saddle-point value of $i\tilde{\lambda} = \Delta^2$ determined through Eq.~(\ref{agap}) can take negative values, provided $\sqrt{|k|^2 - |\Delta^2|}$ is real for all possible values of $k$. 


We now consider the constraint of Eq.~(\ref{evenodd}), requiring that the wavefunctional must be either an even or odd function of the $z_{\alpha}$. These two cases correspond to an even or odd number of spins in the underlying lattice antiferromagnet of interest. In terms of the results in Section \ref{spec}, this means we need to calculate the full spectrum for all of the relevant boundary conditions, and then separate the spectrum into the states with even particle-number states and odd particle-number states to describe the two possibilities. 

\begin{table}
\begin{center}
\begin{tabular}{| c || c | c |}
\hline
Deg. & $\kappa = 0$ & $\kappa = 1$ \\
\hline
1 & 0 & \\
2 & 1.921 & \\
9 & 3.0239 & \\
1 & 3.0244 & \\
25 & 6.048 & \\
66 & 7.111 & 7.111 \\
60 & & 7.975 \\
1 & 8.126 & \\
49 & 9.072 & \\
\hline
\end{tabular}
\caption{Energy splittings $L(E - E_0)$ and their degeneracies at $s = s_c$ for the $O(4)^*$ transition from the large-$N$ expansion with $\tau=i$. Here, $\kappa = L|k|/2\pi$. The ground state energy relative to $L = \infty$ is $L E_0 = -1.317$. Here, we restrict to states that are even in the fields $z_{\alpha}$, which corresponds to an antiferromagnet with an even number of spins.}
\end{center}
\label{sqspecstar}
\end{table}

\begin{table}
\begin{center}
\begin{tabular}{| c || c | c | c | c | c | c |}
\hline
Deg. & $\kappa = 0$ & $\kappa = 1/2$ & $\kappa = 1/\sqrt{2}$ & $\kappa=1$ & $\kappa = \sqrt{5}/2$ & $\kappa = \sqrt{2}$ \\
\hline
4 & 1.512 & & & & & \\
16 &  & 4.516 & & & & \\
16 & 4.536 & & & & & \\
16 &  &  & & 6.463 & & \\
16 & & & 6.694 & & & \\
36 & 7.560 & & & & & \\
32 & & & & & 8.719 & \\
 16 & & & & & & 9.013 \\
\hline
\end{tabular}
\caption{Energy splittings from $L (E - E_0)$ for the $O(4)^*$ transition from the large-$N$ expansion with $\tau=i$ and $N=\infty$. Here, $\kappa = L|k|/2\pi$, and we restrict to states that are odd in the fields $z_{\alpha}$, which corresponds to an antiferromagnet with an odd number of spins. We are measuring the energies with respect to the lowest energy in the O($4$) model, $L E_0 = -1.317$, for comparison with Table \ref{sqspecstar}.}
\label{sqspecodd}
\end{center}
\end{table}

\subsection{Evolution of the spectrum of a function of $s - s_c$}
\label{evotopo}

When considering the deviation from the critical point, the topologically nontrivial sectors correspond to extra features in the two neighboring phases. In a $\mathbb{Z}_2$ spin liquid, the ground state on a torus will exhibit a four-fold degeneracy up to exponential splitting in the system size. In addition, excited states in each topological sector will also contain a four-fold degeneracy corresponding to excitations in the background of different flux sectors through the holes of the torus. This topological degeneracy is the only remnant of the vison particle, which has been integrated out to obtain the O($2N)^\ast$ model, so our theory only captures the spectrum at energies well below the vison mass. 

\subsubsection{Topological phase}

This degeneracy is easily verified in our model; as shown above, the phase with $s > s_c$ will have an energy gap even in an infinite volume, which results in the spectrum showing a weak dependence on boundary conditions. This will cause the different topological sectors to become degenerate up to an exponential splitting of magnitude $e^{-mL}$ where $m = 2 \pi (s - s_c)$. From solving Eq.~(\ref{agap}) for $s \gg s_c$, one find that in all four sectors the gap approaches $\Delta = m$ up to exponential corrections in the system size, and similarly the ground state energies in this limit will become exponentially close.

\subsubsection{Magnetically ordered phase}
\begin{figure}
\centering
\includegraphics[width=17cm]{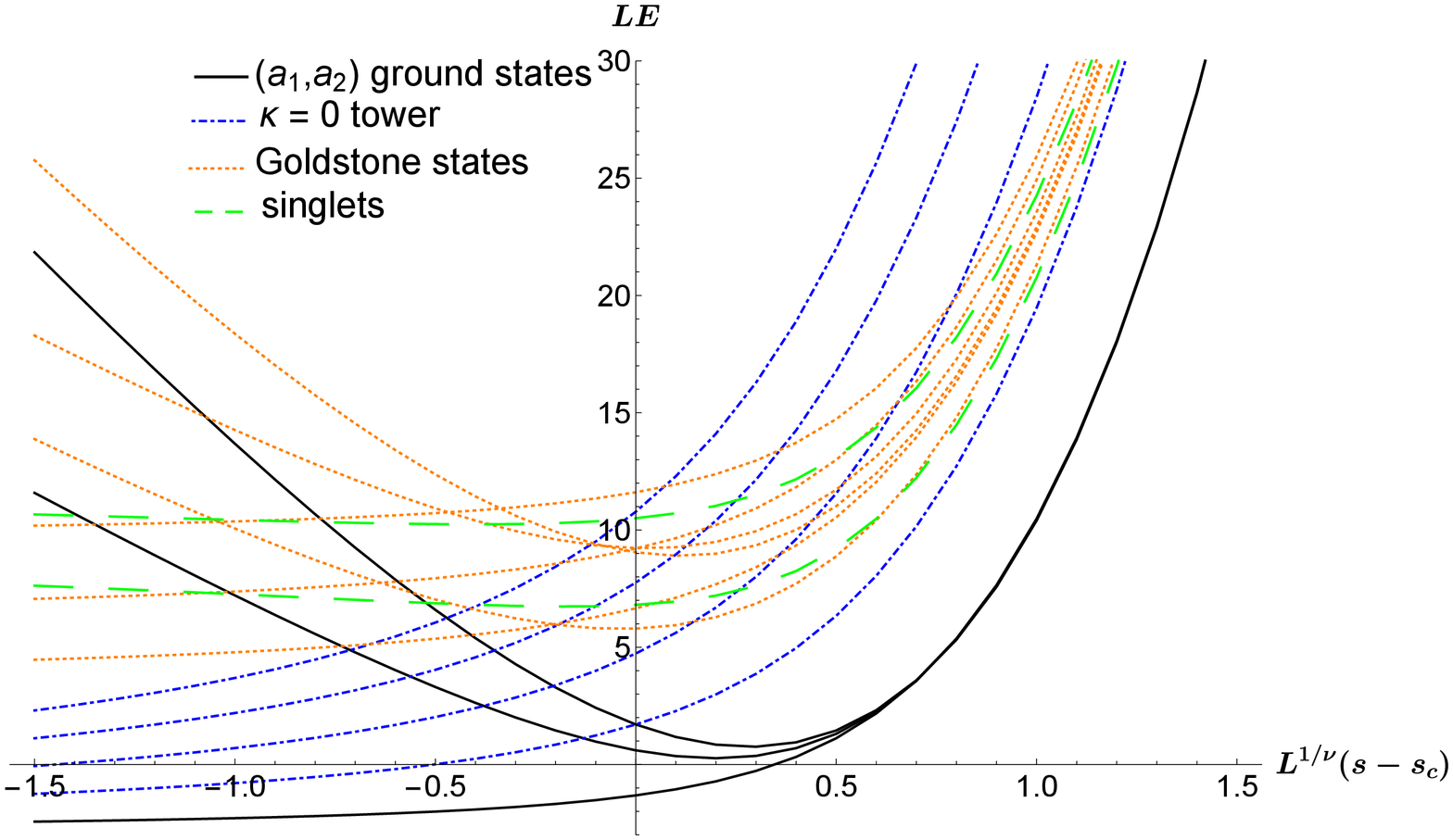}
\caption{(Color online) The evolution of the spectrum $LE$ for the O($4)^\ast$ model as a function of the tuning parameter $L^{1/\nu}(s - s_c)$ on the square torus, $\tau = i$. Note that $\nu = 1$ to leading order in $1/N$. The energy levels are defined so that $E = 0$ at $s = s_c$ and $L = \infty$. We label the states by their behavior in the ordered region, distinguishing between the tower, the Goldstone modes, and the singlet states. We also distinguish the four ``ground states'' of the different sectors ($a_1,a_2$) according to Table \ref{adef}, though the (A,P) and (P,A) sectors are degenerate for the square geometry. These states become degenerate in the topological phase, while they represent $\mathbb{Z}_2$ vortices in the magnetic phase. Our choice of states is not exhaustive, but highlights the main features of the proximate phases.}
\label{o4starplot}
\end{figure}
In the magnetically ordered phase, $s < s_c$, the antiperiodic boundary conditions have an interpretation as vortices of the order parameter. This can be seen from the parametrization of the order parameter in terms of the spinon degrees of freedom in Eq.~(\ref{spinors}). As the spinon field undergoes a smooth non-contractible twist around a cycle of the torus, $z_{\alpha} \rightarrow -z_{\alpha}$, the physical order parameter returns to its original configuration after traversing a topologically nontrivial path in order parameter space. These correspond to vortices associated with the first homotopy group, $\pi_1(SO(3)) = \mathbb{Z}_2$. Note that by only allowing twists in the order parameter around the torus, we are ignoring local vortex configurations. This simplification is analogous to ignoring the local vison excitations in the spin liquid phase, since a local vortex will have some extra energy cost due to its core.

The energy cost of a vortex can be estimated by dimensional analysis. On general grounds, in the ordered phase we can write the energy functional for the phase $\theta(x)$ of the order parameter as
\beq
\mathcal{E} = \frac{\rho_s}{2} \int d^2x \left( \nabla \theta_{\alpha} \right)^2
\eeq
where $\rho_s$ is a ``spin stiffness'' (really the stiffness of the condensed $z_{\alpha}$ fields rather than the underlying spin order parameter), given by $\rho_s \sim N (s_c - s)$ close to the large-$N$ critical point \cite{CSS93,CSS94}. We consider a smooth configuration of the field from $z_{\alpha} \rightarrow -z_{\alpha}$ as the order parameter winds around either cycle, which have lengths $|\omega_{1,2}|$. This contributes a gradient of order $ \nabla z_{\alpha} \sim 1/|\omega_{1,2}|$, and the energy cost will be
\beq
\mathcal{E} \sim N (s_c - s) \frac{\mathcal{A}}{|\omega_{1,2}|^2}.
\eeq
The estimate can be checked against the current model. For $s \ll s_c$, the solution of of gap equation becomes
\beq
\Delta^2 = \frac{1}{\mathcal{A}^2 (s_c - s)^2} - |k_{\mathrm{min}}|^2
\eeq
where $|k_{\mathrm{min}}|$ is the minimum value of $|k|$ allowed in a given topological sector (so $k_{\mathrm{min}}$ is always zero in the (P,P) sector). Solving Eq.~(\ref{aenergy}) for the energy of a vortex in this limit gives
\beq
E_{\mathrm{vortex}} \equiv E_{0} - E_{0,(P,P)} = \frac{N \mathcal{A} (s_c - s)}{2}|k_{\mathrm{min}}|^2 \qquad s \ll s_c
\eeq
This agrees with the above estimate since $|k_{\mathrm{min}}|^2 \sim 1/|\omega_{1,2}|^2$ in the different sectors.

\subsection{Results}
\begin{figure}
\centering
\includegraphics[width=8.73cm]{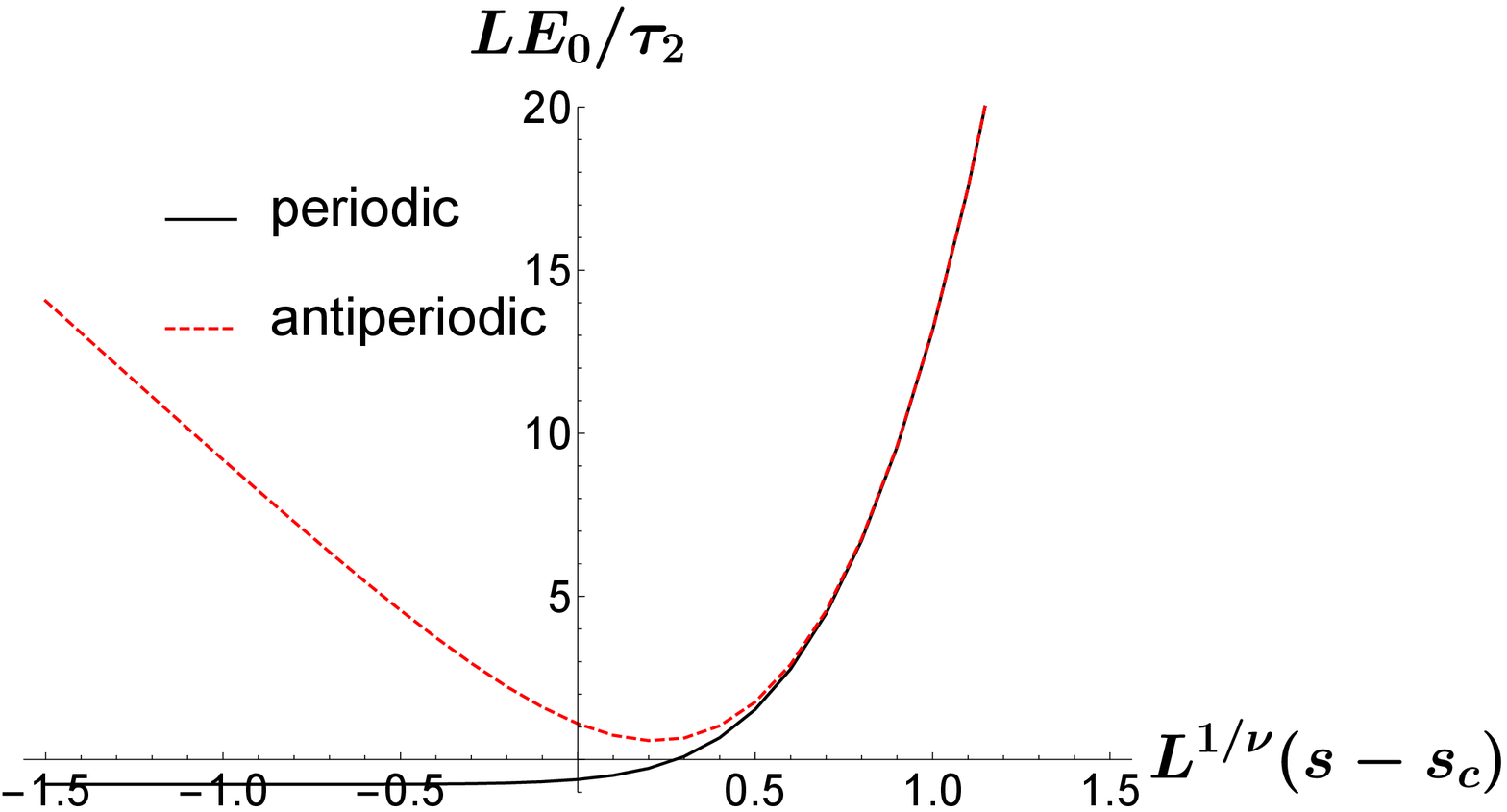}
\includegraphics[width=8.73cm]{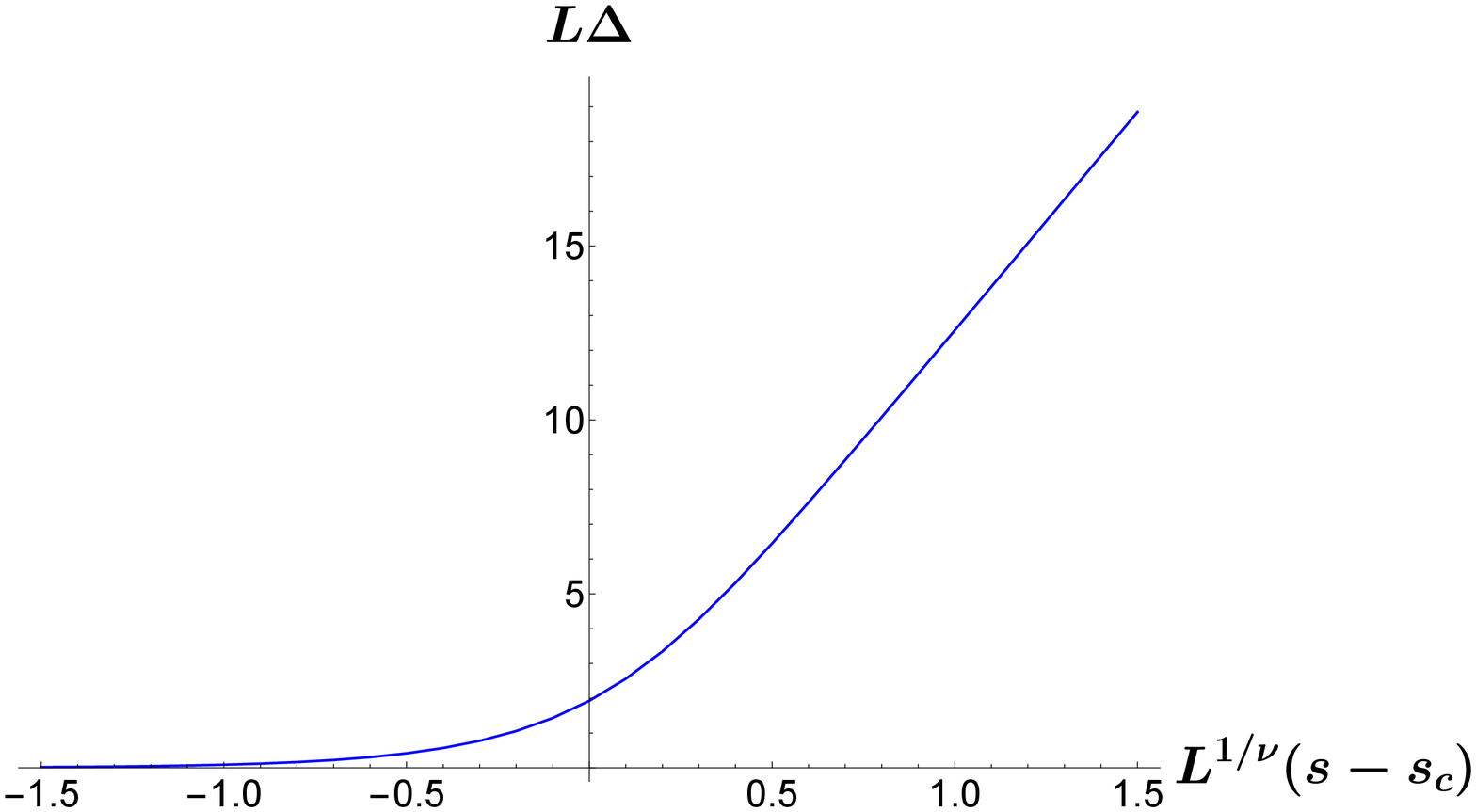}
\caption{(Color online) Left: The splitting between the energy densities of the periodic and antiperiodic sectors for the O(4)$^\ast$ model on the infinite cylinder. The energy levels are defined so that $E_{0}/\mathcal{A} = 0$ at $s = s_c$ and $L = \infty$. Right: The energy gap above the ground state for the O(4)$^\ast$ model on the cylinder as a function of $s - s_c$. The spectrum above this gap is continuous. }
\label{cylindst}
\end{figure}
We give the results for the low-lying O($4)^\ast$ spectrum on a square torus at criticality in Tables \ref{sqspecstar} and \ref{sqspecodd}, which contain the even and odd spin results respectively. We also give the evolution of the spectrum as a function of $s - s_c$ in Figure \ref{o4starplot}, choosing some representative states to depict the nature of the two phases. We also give universal results for the cylindrical limit in Figure \ref{cylindst}. We plot the splitting between the ground state energy densities in the two sectors, as well as the excitation gap which is simply twice the gap for the O(4) model. Above the excitation gap, the spectrum becomes a continuum due to the momentum along the infinite direction, so there are no universal energy levels.

\begin{figure}
\centering
\includegraphics[width=17cm]{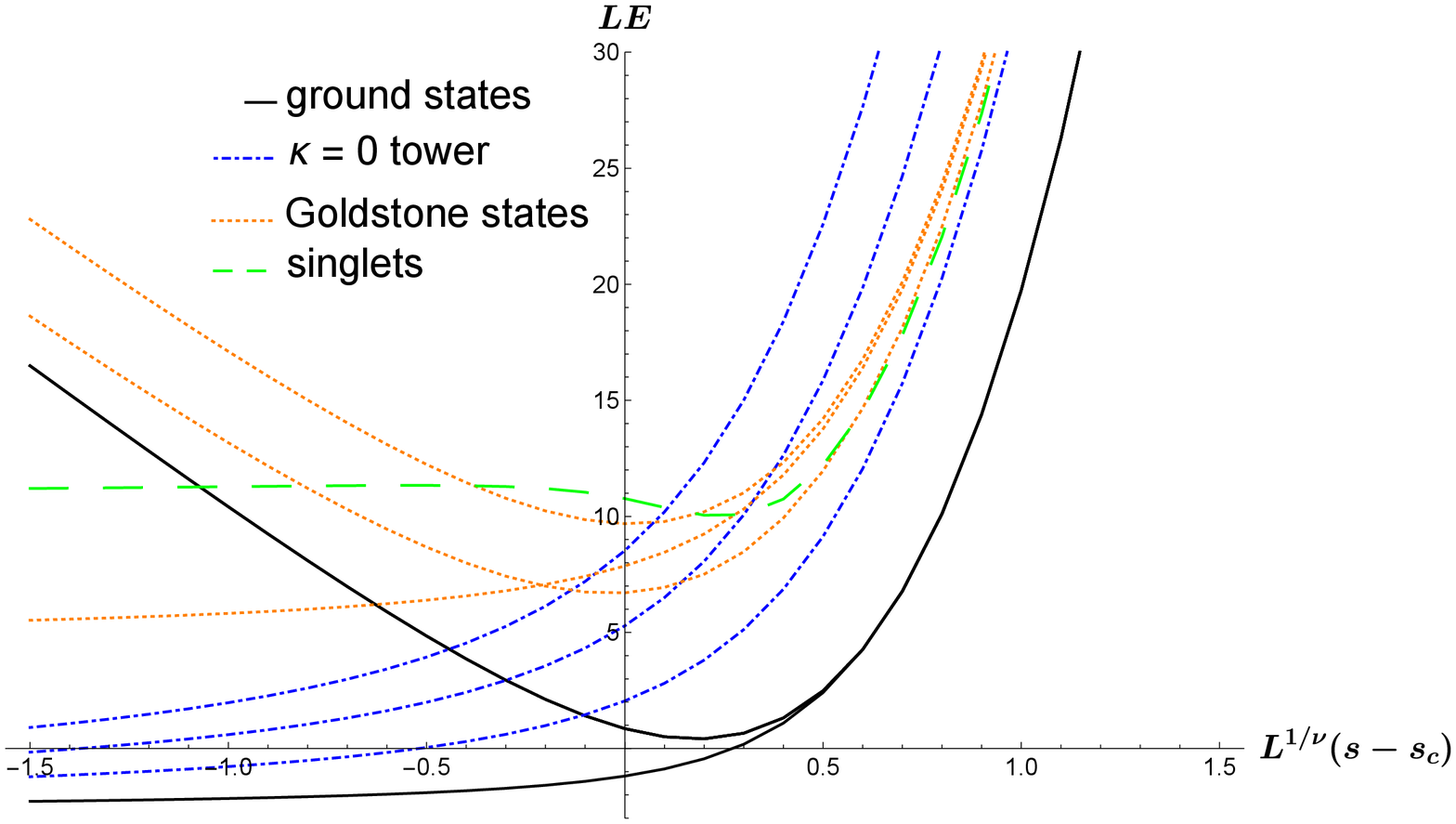}
\caption{(Color online) The evolution of the spectrum $LE$ for the O($4)^\ast$ model as a function of the tuning parameter $L^{1/\nu}(s - s_c)$ on the triangular torus, $\tau = e^{i \pi/3}$. Note that $\nu = 1$ to leading order in $1/N$. The energy levels are defined so that $E = 0$ at $s = s_c$ and $L = \infty$. We label the states by their behavior in the ordered region, distinguishing between the tower, the Goldstone modes, and the singlet states. Note that the three sectors (A,P), (P,A), and (A,A) are degenerate in this geometry. Our choice of states is not exhaustive, but highlights the main features of the proximate phases.}
\label{trio4starplot}
\end{figure}

We also comment on the triangular torus, $\tau = e^{i \pi/3}$. This is an interesting case because numerical simulations on the triangular lattice are more easily performed using this boundary condition, so these results have relevance to future studies on the $J_1$-$J_2$ Heisenberg model where the antiferromagnetic-spin liquid transition has been reported. For this special value of the modular parameter, it turns out that all three nontrivial topological sectors are exactly degenerate. This is due to the choice $e^{i \pi/3}$ being invariant under the modular transformation $\tau \rightarrow -1/(\tau-1)$, see the discussion below Eq.~(\ref{modular}). In addition, this torus has a discrete six-fold rotational symmetry, resulting in a highly degenerate spectrum for finite-momentum states. The evolution of the spectrum for the triangular torus is shown in Figure \ref{trio4starplot}.

\section{Anisotropic corrections}
\label{aniso}
We now consider to the leading irrelevant operator in our theory,
\beq
\mathcal{L}' = \gamma |z_\alpha^\ast \nabla z_\alpha|^2. \label{irrel}
\eeq
Asymptotically close to the critical point, this term is irrelevant and will not contribute to universal physics. However, this term is \emph{dangerously} irrelevant because it breaks the O($2N$) symmetry down to SU($N$) for any deviation from the scaling limit. Therefore, the actual energy levels for the transition will organize into SU($N$) multiplets for any lattice model, with a splitting determined by $\gamma$. The coefficient $\gamma$ is non-universal and will be determined by microscopics, so in principle one must fit its value to a given spectrum. 

We begin by discussing the nature of the splitting in terms of representation theory. The real and imaginary parts of $z_{\alpha}$ transform together as an O($2N$) vector, but this representation will transform reducibly under the SU($N$) symmetry of Eq.~(\ref{irrel}). Labelling the irreducible representations by their dimension, the splitting of the O($2N$) vector into SU($N$) representations is
\beq
2N \longrightarrow N \oplus \overline{N}, \label{embed}
\eeq
where $N$ and $\overline{N}$ are the fundamental and anti-fundamental representations of SU($N$), which we will shortly associate with spinons and anti-spinons. We can analyze the breaking of higher representations of O($2N$) by taking tensor products of the fundamental representation. For example, the splitting of the two-particle states can be obtained by taking the antisymmetric or symmetric tensor product of the O($2N$) vector, and use the known properties for adding SU($N$) representations
\bea
\left[ \left( N \oplus \overline{N} \right) \otimes \left( N \oplus \overline{N} \right)  \right]_A &=& \frac{N(N-1)}{2}\oplus \frac{\overline{N(N-1)}}{2} \oplus \left(N^2-1\right) \oplus 1 \nn
\left[ \left( N \oplus \overline{N} \right) \otimes \left( N \oplus \overline{N} \right)  \right]_S &=& \frac{N(N+1)}{2}\oplus \frac{\overline{N(N+1)}}{2} \oplus \left(N^2-1\right) \oplus 1 \label{2part1}
\eea
where the subscripts indicate antisymmetrizing or symmetrizing the direct product with respect to the ordering of the O($2N$) indices. Since the symmetric representation of O($2N$) contains an irreducible singlet, it must coincide with the singlet state in the last line of Eq.~(\ref{2part1}).

We can make contact with our expressions in Section \ref{spec} by defining spinon and anti-spinon operators and relating them to the O($2N$) vector operators $b_{\alpha}^{\dagger}$. We expand the $z_{\alpha}$ field as
\beq
z_{\alpha} = \frac{1}{\mathcal{A}^{1/2}} \sum_{k \neq 0} \frac{e^{ik \cdot x}}{\sqrt{2 E_1(k)}} \left( a_{\alpha}(k) + c_{\alpha}^{\dagger}(-k) \right).
\eeq
Here, the dot product is given by $k \cdot x \equiv \mathrm{Re}(k x^{\ast})$, and $E_1(k) = \sqrt{|k|^2 + \Delta^2}$ is the single-particle energy at $N=\infty$. Here, we are assuming that the perturbation $\gamma$ does not shift the saddle-point value of the path integral, so we can perturb around the $N=\infty$ spectrum. Since $z_{\alpha}$ transforms as an SU($N$) vector, the particles created by $c^{\dagger}_{\alpha}$ are spinons and the particles created by $a^{\dagger}_{\alpha}$ are anti-spinons. We can identify these with the O($2N$) bosons defined earlier
\bea
c^{\dagger}_{\alpha} &=& \frac{1}{\sqrt{2}} \left( b^{\dagger}_{\alpha} + i b^{\dagger}_{\alpha + N} \right) \nn
a^{\dagger}_{\alpha} &=& \frac{1}{\sqrt{2}} \left( b^{\dagger}_{\alpha} - i b^{\dagger}_{\alpha + N} \right).
\eea
From these relations it is straight-forward to check that the embedding in Eq.~(\ref{embed}) holds. The decomposition of the two-particle states can be written
\bea
b^{\dagger}_{[\alpha} b^{\dagger}_{\beta]} &\longrightarrow& c^{\dagger}_{[\alpha} c^{\dagger}_{\beta]} + a^{\dagger}_{[\alpha} a^{\dagger}_{\beta]} + \left( c^{\dagger}_{\alpha}a^{\dagger}_{\beta} - a^{\dagger}_{\beta}c^{\dagger}_{\alpha} - \frac{\delta_{\alpha \beta}}{N}\left( c^{\dagger}_{\gamma}a^{\dagger}_{\gamma} - a^{\dagger}_{\gamma}c^{\dagger}_{\gamma} \right) \right) + \frac{\delta_{\alpha \beta}}{N}\left( c^{\dagger}_{\gamma}a^{\dagger}_{\gamma} - a^{\dagger}_{\gamma}c^{\dagger}_{\gamma} \right) \nn
b^{\dagger}_{(\alpha} b^{\dagger}_{\beta)} &\longrightarrow& c^{\dagger}_{(\alpha} c^{\dagger}_{\beta)} + a^{\dagger}_{(\alpha} a^{\dagger}_{\beta)} + \left( c^{\dagger}_{\alpha}a^{\dagger}_{\beta} + a^{\dagger}_{\beta}c^{\dagger}_{\alpha} - \frac{\delta_{\alpha \beta}}{N}\left( c^{\dagger}_{\gamma}a^{\dagger}_{\gamma} + a^{\dagger}_{\gamma}c^{\dagger}_{\gamma} \right) \right) \nn
b^{\dagger}_{\gamma} b^{\dagger}_{\gamma} &\longrightarrow& c^{\dagger}_{\gamma}a^{\dagger}_{\gamma} + a^{\dagger}_{\gamma}c^{\dagger}_{\gamma} \label{sunirreps}
\eea
where the indices on the left run to $2N$ while the indices on the right run to $N$. Once again, if the two states carry the same momentum there is no antisymmetric contribution.

We now apply perturbation theory on the degenerate states, using Eq.~(\ref{sunirreps}) to diagonalize the perturbation. We define the dimensionless coupling $\tilde{\gamma} \equiv \gamma/L$ as well as the shorthand $\chi_{\alpha}(k) \equiv a_{\alpha}(k) + c_{\alpha}^{\dagger}(-k)$, and obtain the interaction Hamiltonian
\bea
V_{\gamma} = \frac{\tilde{\gamma}}{\tau_2 L} \sum_{k_1,k_2,k_3 \neq 0} \frac{k_2 \cdot k_3}{4 \sqrt{ E_1(k_1) E_1(k_2) E_1(k_3) E_1(k_1-k_2+k_3)}} \chi^{\dagger}_{\alpha}(k_1)\chi_{\alpha}(k_2)\chi^{\dagger}_{\beta}(k_3)\chi_{\beta}(k_1-k_2+k_3). \qquad \label{gammaint}
\eea
The single particle energies of spinons and anti-spinons are shifted by the same amount, so there is no splitting to one-particle states to leading order.

We will explicitly compute the shift in energies for the two-particle states in Eq.~(\ref{sunirreps}), which are all degenerate at $N=\infty$ except for the singlet state in the last line. The perturbation will split these states, and can also split any possible degeneracy between states with the same total momentum. We first ignore the latter possibility, which does not occur for any of the states listed in the above tables. Recall that the two-particle state energies can be written as $E_2(k) = E_1 (q) + E_1 (k-q)$ for some value of $q$. Then the splitting of the antisymmetric representation is
\bea
\frac{N(N-1)}{2}, \frac{\overline{N(N-1)}}{2}: &\qquad& \Delta E_{\mathrm{asym}}(k) = - \frac{\tilde{\gamma}}{\tau_2 L} \frac{|k - 2 q|^2}{4 E_1(q) E_1(k - q)} \nn
N^2 - 1: &\qquad& \Delta E_{\mathrm{adj}}(k) = - \frac{\tilde{\gamma}}{\tau_2 L} \frac{2 q \cdot \left(k - q \right)}{4 E_1(q) E_1(k - q)} \nn
1: &\qquad& \Delta E_{\mathrm{s}}(k) = \frac{\tilde{\gamma}}{\tau_2 L} \frac{N \left( |q|^2 + |k - q|^2 \right) - 2 q \cdot (k-q)}{4 E_1(q) E_1(k - q)},
\eea
while for the symmetric representation,
\bea
\frac{N(N+1)}{2}, \frac{\overline{N(N+1)}}{2}: &\qquad& \Delta E_{\mathrm{sym}}(k) = \frac{\tilde{\gamma}}{\tau_2 L} \frac{|k|^2}{4 E_1(q) E_1(k - q)} \nn
N^2 - 1: &\qquad& \Delta E_{\mathrm{adj}}(k) = - \frac{\tilde{\gamma}}{\tau_2 L} \frac{2 q \cdot \left( k - q \right)}{4 E_1(q) E_1(k - q)}.
\eea
The subscripts refer to the states being in the symmetric, antisymmetric, singlet, or adjoint representations of SU($N$).

Summarizing the results to first order in $\gamma$, the degeneracy of the antisymmetric representation breaks down from $N(2N-1)$ to $N(N-1)$, $N^2 - 1$, and $1$, while the degeneracy of the symmetric traceless tensor representation breaks down from $(2N-1)(2N+2)/2$ to $N(N+1)$ and $N^2 - 1$. 

Note that the first-order correction is zero if the unperturbed particles all have zero momentum. Therefore, to first order there is no splitting of the ``tower of states'' in the antiferromagnetic phase. Although we do not compute the magnitude for the splitting of the states in the tower, we comment on the expected representations which should appear. In Section \ref{secevo} we saw that the tower of states for the O($2N$) model all belong to the symmetric traceless tensor representations. For the case of interest, $N=2$, the allowed degeneracies in the tower becomes $(2\ell + 1)^2$ for $\ell = 0,1, 2, ...$ where we use the constraint that only an even number of particles are allowed. Repeating the above analysis by forming symmetric products and subtracting out the traces, one finds that each of these states decomposes into $(2 \ell + 1)$ different SU(2) representations each with spin-$\ell$. We also note that the spacing of the even-particle spectrum for the O($4)^\ast$ model should be proportional to $2\ell (2 \ell + 2) \propto \ell (\ell + 1)$, which agrees with the spacing for the tower in an SU(2) antiferromagnet \cite{Azaria93}. This qualitative structure of the spectrum, with $(2 \ell + 1)$ inequivalent spin-$\ell$ multiplets in the tower becoming approximately degenerate close to the critical point, is an interesting feature of this theory which could give good evidence for the existence of an O($4)^\ast$ transition and a neighboring spin liquid phase.

\begin{table}
\begin{tabular}{| c | c | c | c || c | c |}
\hline
deg. at $\gamma=0$ & $\sqrt{\mathcal{A}}E_2$ & $\kappa$ & $\tilde{q}$ & deg. at $\mathcal{O}(\gamma)$ & $\sqrt{\mathcal{A}} \Delta E$ \\
\hline
9 & 3.0239 & 0 & 0 & 9 & 0 \\
\hline
\multirow{4}{*}{30} & \multirow{4}{*}{7.111} & \multirow{4}{*}{0} & \multirow{4}{*}{1/2} & 4 & $-1.47\tilde{\gamma}$ \\
 & & & & 12 & 0 \\
  & & & & 12 & 0.73$\tilde{\gamma}$ \\
   & & & & 2 & $1.47\tilde{\gamma}$ \\
   \hline
\multirow{2}{*}{36} & \multirow{2}{*}{7.111} & \multirow{2}{*}{1} & \multirow{2}{*}{1/2} & 12 & $-0.73\tilde{\gamma}$ \\
 & & & & 24 & 0 \\
 \hline
\multirow{4}{*}{60} & \multirow{4}{*}{7.975} & \multirow{4}{*}{1} & \multirow{4}{*}{0} & 8 & $-1.01\tilde{\gamma}$ \\
& & & & 24 & 0 \\
& & & & 24 & 1.01$\tilde{\gamma}$ \\
& & & & 4 & 2.02$\tilde{\gamma}$ \\
\hline
\end{tabular}
\caption{The two-particle states in the even sector of the critical O($4)^\ast$ spectrum, taken from Table \ref{sqspecstar}, and their splitting due to the perturbation. The energies of these states are written as $E_2(k) = E_{gs} + E_1 (q) + E_1 (k-q)$, and we list the scaled momenta, $\kappa = L |k|/2\pi$ and $\tilde{q} = L|q|/2\pi$. For further details, see the text.}
\label{splittab}
\end{table}

For a definite example, we revisit the results for the even sector of the O($4)^\ast$ model on the square torus. In Table \ref{splittab}, we explicitly show all the two-particle states from Table \ref{sqspecstar} which are split by the perturbation, and give the magnitude of the splitting. Note that the numerical value of all energies will be shifted from their unperturbed values, but here we only give the energy splitting between states. The states listed in Table \ref{splittab} turn out to be the only states in Table \ref{sqspecstar} which are split at first-order in $\gamma$.

In principle, one can continue this process to higher-particle states, and to higher order in $\gamma$. For a more complex O($2N$) multiplet, one finds how the SU($N$) representations fit inside the larger group, and use this to diagonalize the perturbation within the degenerate multiplet. 

\section{Conclusions}
\label{sec:conc}

There have been extensive discussions in the literature on the nature of the finite size and low energy spectrum of quantum antiferromagnets in antiferromagnetically ordered and gapped topological phases. For magnetically-ordered antiferromagnets, we have 
the well-known ``tower of states'' \cite{PWA52,Siggia89,Bernu92,Bernu94,Azaria93,White07} obtained 
from the excitations of a quantum rotor representing the spatially uniform collective quantum fluctuations of all the spins; such a spectrum is characteristic signature of the spontaneously broken spin rotation
symmetry. On the other hand, antiferromagnets with an energy gap and topological order have low energy states whose
energy differences are exponentially small in the system size; again, this nearly degenerate spectrum is a characteristic
signature of the topological order in this phase of the antiferromagnet. 

In the present paper, we have presented results on the evolution of the spectrum between the above two limits.
We examined a two-dimensional antiferromagnet, with global SU(2) spin rotation symmetry, which undergoes a transition between a gapped $\mathbb{Z}_2$ spin liquid 
and coplanar antiferromagnetic order. Such a transition is described by a O$(4)^\ast$ conformal field theory in 2+1 dimensions, 
which is closely related to the O(4) Wilson-Fisher conformal field theory. We showed that the quantum critical point
has a universal spectrum, in which the energy levels are universal numbers times $1/L$, where $L$ is the spatial system size.
This spectrum contains features which descend from the phases found on either side of the critical point.
The topological degeneracy on the gapped side evolves into non-trivial boundary conditions and selection rules on 
the operators of the conformal field theory. And the spontaneously broken spin-rotation symmetry on the other side
yields low-lying states with non-zero spin at the critical point. 

We hope that our results will aid in analyzing numerical data on lattice quantum antiferromagnets which
undergo transitions from antiferromagnetically ordered to spin liquid states. With the available data on the manner
in which the ``tower of states'' evolve into the spin liquid across a quantum critical point, strong constraints become
available on identifying the topological order in the spin liquid.

\section*{Acknowledgments} 
We thank A. Paramekanti and A. Thomson for helpful discussions, and L.-P. Henry, A.~La\"uchli, and M.~Schuler for collaboration on a related
project \cite{HLSSW16}.
This research was supported by the NSF under Grant DMR-1360789.
Research at Perimeter Institute is supported by the Government of Canada through Industry Canada 
and by the Province of Ontario through the Ministry of Research and Innovation.   

\appendix
\section{Loop sums}
\label{dimreg}

Here we review the calculation of loop diagrams in a finite volume using dimensional regularization. 

To remind the reader of our notation, we parametrize the coordinates on the spatial torus in complex coordinates $w = x + iy$, and denote the two periods in these coordinates as $\omega_1$ and $\omega_2$ (see Fig. \ref{torusfig}). We define the modular parameter $\tau \equiv \omega_2/\omega_1$ and the length scale $L \equiv |\omega_1|$. In this geometry, the basis vectors of the dual lattice are given by
\beq
k_1 = -i\omega_2/\mathcal{A}, \qquad k_2 = i \omega_1/\mathcal{A}.
\eeq
The eigenvalues of the Laplacian are dependent on the boundary conditions of the torus. We consider the fields to be either periodic or anti-periodic in either direction. With this in mind, we write the eigenvalues of the Laplacian as
\beq
|k_{n,m}|^2 = \left( 2 \pi \right)^2 \left| (n + a_1) k_1 + (m + a_2) k_2 \right|^2, \qquad n, m \in \mathbb{Z}.
\label{kdef}
\eeq
Here, the numbers $a_1$ and $a_2$ parametrize whether the boundary conditions on the fields are periodic or anti-periodic in the directions $\omega_1$ and $\omega_2$, see Table \ref{adef}.

A general one-loop diagram will be of the form
\beq
\sum_{n,m \in \mathbb{Z}} \frac{1}{\left( |k_{n,m}|^2 + \Delta^2 \right)^s} = \left( \frac{ \tau_2 L}{2 \pi} \right)^{2s} \sum_{n,m \in \mathbb{Z}} \frac{1}{\left( |m + a_2 + (n + a_1)\tau |^2 + \gamma^2 \right)^s} \label{1loop}
\eeq
where $\tau = \tau_1 + i \tau_2$ and $\gamma = \tau_2L \Delta/2 \pi$ (we have used $\tau_2 = \mathcal{A}/L^2$).

We now generalize this sum to arbitrary dimension. This is done by promoting the two-dimensional vector $(n + a_1,m + a_2)$ to a $d$-dimensional vector of (half) integers for the (anti-)periodic case. Then in (\ref{1loop}) we simply take the sums to be over $n, m \in \mathbb{Z}^{d/2}$. We will write the sums as
\beq
g_{s}^{(d)}(\Delta,\tau) = \sum_{n,m \in \mathbb{Z}^{d/2}} \frac{1}{\left( |m + a_2 + (n + a_1)\tau |^2 + \gamma^2 \right)^s}. \label{gdef}
\eeq

The summand is rewritten using the identity
\beq
\frac{1}{A^s} = \frac{\pi^s}{\Gamma(s)} \int_0^{\infty} d\lambda \lambda^{s-1} e^{- \pi \lambda A}
\eeq
giving
\beq
g_{s}^{(d)} = \frac{\pi^s}{\Gamma(s)} \int_0^{\infty} d\lambda \lambda^{s-1} e^{- \pi \lambda \gamma^2} \sum_{n,m \in \mathbb{Z}^{d/2}} \exp \left(- \pi \lambda |m + a_2 + (n + a_1)\tau |^2\right).
\eeq
We can now write the sum in terms of the two-dimensional Riemann theta function, defined as
\beq
\Theta\left(\lambda,\mathbf{\Omega},\mathbf{u}\right) \equiv \sum_{\mathbf{n} \in \mathbb{Z}^2} \exp\big(-\pi \lambda \mathbf{n}^{\intercal} \cdot \mathbf{\Omega} \cdot \mathbf{n} - 2 \pi \mathbf{n}^{T} \cdot \mathbf{u} \big) \label{rtheta}
\eeq
where $\Omega$ is a $2\times 2$ matrix and $\mathbf{u}$ is a two-dimensional vector. Then
\beq
g_s^{(d)} = \frac{\pi^s}{\Gamma(s)} \int_0^{\infty} d\lambda \lambda^{s-1} e^{-\pi \lambda \gamma^2} \exp\left(- \pi \lambda \gamma^2 - \frac{d \pi \lambda}{2} \left( (a_1 \tau_2)^2 + (a_2 + a_1 \tau_1)^2 \right) \right) \Theta\left(\lambda,\mathbf{\Omega}(\tau),\mathbf{v}_1\right)^{d/2} \label{unreg}
\eeq
where
\beq
\mathbf{\Omega}(\tau) = 
\begin{pmatrix}
| \tau |^2 & \tau_1 \\
\tau_1 & 1 
\end{pmatrix}, \qquad \mathbf{v}_1 = \lambda \begin{pmatrix} \tau_1\left( a_2 + a_1 \tau_1 \right) + a_1 \tau_2^2 \\ a_2 + a_1 \tau_1 \end{pmatrix}.
\eeq

As with the original sum, the function (\ref{unreg}) converges whenever $s > d/2$, while for $s < d/2$, the integral diverges for small values of $\lambda$. We proceed by splitting the integral into two parts, $\int_0^{\infty} = \int_0^1 + \int_1^{\infty}$, and working on the divergent piece. Using the mathematical identity
\beq
\Theta\left(\lambda,\mathbf{\Omega},\mathbf{u}\right) = \frac{1}{\lambda \sqrt{\det \mathbf{\Omega}}} \exp \left( \frac{\pi}{\lambda} \mathbf{u}^T \cdot \mathbf{\Omega}^{-1} \cdot \mathbf{u} \right) \Theta\left(\frac{1}{\lambda},\mathbf{\Omega}^{-1},-\frac{i}{\lambda}\mathbf{\Omega}^{-1} \cdot\mathbf{u} \right),
\eeq
the integral at small $\lambda$ becomes
\bea
\tau_2^{-d/2}\frac{\pi^s}{\Gamma(s)} \int_0^{1} d\lambda \lambda^{s-1-d/2} e^{- \pi \lambda \gamma^2} \Theta\left(\frac{1}{\lambda},\mathbf{\Omega}(\tau)^{-1},\mathbf{v}_2\right)^{d/2} \nn
= \tau_2^{-d/2}\frac{\pi^s}{\Gamma(s)} \int_1^{\infty} d\lambda \lambda^{d/2-s-1} e^{- \pi \gamma^2/\lambda} \Theta\left(\lambda,\mathbf{\Omega}(\tau)^{-1},\mathbf{v}_2\right)^{d/2}
\eea
with $\mathbf{v}_2 = -i(a_1,a_2)$. Since $\Theta \rightarrow 1$ for large $\lambda$, we see that the integral has the expected UV divergence. In this paper, we evaluate sums with $s = 1/2$ and $s = -1/2$ in $d=2$, so we add and subtract the divergent terms for these cases, evaluating integrals where possible in the convergent region $s>d/2$:
\bea
\tau_2^{-d/2}\frac{\pi^s}{\Gamma(s)} \int_1^{\infty} d\lambda \lambda^{d/2-s-1} \left( e^{- \pi \gamma^2/\lambda} \Theta\left(\lambda,\mathbf{\Omega}(\tau)^{-1},\mathbf{v}_2\right)^{d/2} - 1 - \frac{\pi \gamma^2}{\lambda} \right) \nn
+ \tau_2^{-d/2}\frac{\pi^s}{\Gamma(s)} \left(\frac{1}{s-d/2} + \frac{\pi \gamma^2}{1 + s - d/2} \right).
\eea
Now that the integrals which were only convergent for $s>d/2$ have been evaluated, we analytically continue the result to the dimensionality of interest, and $g_s^{(d)}$ will only have simple poles on the complex plane. 

To summarize, we have
\bea
&& \qquad \qquad \qquad \qquad \qquad \quad \sum_{k} \frac{1}{\left( |k|^2 + \Delta^2 \right)^s} = \left( \frac{ \tau_2 L}{2 \pi} \right)^{2s} g_s^{(d)}(\Delta,\tau), \nn
&& g_s^{(d)}(\Delta,\tau) = \frac{\pi^s}{\Gamma(s)} \left\{ \int_1^{\infty} d\lambda \lambda^{s-1} \exp\left(- \frac{\lambda \tau_2^2 L^2 \Delta^2}{4 \pi} - \frac{d \pi \lambda}{2} \left( (a_1 \tau_2)^2 + (a_2 + a_1 \tau_1)^2 \right) \right) \Theta\left(\lambda,\mathbf{\Omega}(\tau),\mathbf{v}_1\right)^{d/2} \right. \nn
&& \qquad \qquad \qquad \qquad + \ \tau_2^{-d/2} \int_1^{\infty} d\lambda \lambda^{d/2-s-1} \left( \exp\left( - \frac{\tau_2^2 L^2 \Delta^2}{4 \pi \lambda} \right) \Theta\left(\lambda,\mathbf{\Omega}(\tau)^{-1},\mathbf{v}_2\right)^{d/2} - 1 + \frac{\tau_2^2 L^2 \Delta^2}{4 \pi \lambda} \right) \nn
&& \qquad \qquad \qquad \qquad + \ \frac{\tau_2^{-d/2}}{s-d/2} - \frac{L^2 \Delta^2}{4 \pi} \frac{\tau_2^{2-d/2}}{1 + s - d/2} \Big\},
\label{specfunc}
\eea
where the Riemann theta function $\Theta$ was defined in (\ref{rtheta}), and we've also defined
\bea
&\mathbf{\Omega}(\tau) = 
\begin{pmatrix}
| \tau |^2 & \tau_1 \\
\tau_1 & 1 
\end{pmatrix}, \qquad
&\mathbf{\Omega}(\tau)^{-1} = 
\frac{1}{\tau_2^2}\begin{pmatrix}
1 & -\tau_1 \\
-\tau_1 & | \tau |^2
\end{pmatrix} \\
&\mathbf{v}_1 = \lambda \begin{pmatrix} \tau_1\left( a_2 + a_1 \tau_1 \right) + a_1 \tau_2^2 \\ a_2 + a_1 \tau_1 \end{pmatrix}, \qquad
&\mathbf{v}_2 = - i \begin{pmatrix} a_1 \\ a_2 
\end{pmatrix}.
\eea

At this point, we note the properties of these sums under modular transformations. Modular transformations are discrete diffeomorphisms on the torus, so we need the spectrum to be invariant under the modular group. This group is generated by the two transformations \cite{DMS97}
\bea
\mathcal{T}: \qquad \tau &\rightarrow& \tau + 1 \nn
\mathcal{S}: \qquad\tau &\rightarrow& -\frac{1}{\tau}.
\eea
Under these transformations, the area $\tau_2 L^2$ is left unchanged. To see how our loop sums transform under these operations, we look at how $|k_{n,m}|^2$ in Eq.~(\ref{kdef}) transforms, since all sums involve some power of this object summed over all integers. A quick calculation finds
\bea
\frac{1}{(2 \pi)^2}|k_{n,m}|^2 &=& \frac{1}{\tau_2 \mathcal{A}} \left[ (n + a_1)^2 (\tau_1^2 + \tau_2^2) + (m + a_2)^2  + 2 (n + a_1)(m + a_2) \tau_1 \right] \nn
&\xrightarrow{\mathcal{T}}& \frac{1}{\tau_2 \mathcal{A}} \left[ (n + a_1 + m + a_2)^2 (\tau_1^2 + \tau_2^2) + (m + a_2)^2  + 2 (n + a_1 + m + a_2)(m + a_2) \tau_1 \right] \nn
&\xrightarrow{\mathcal{S}}& \frac{1}{\tau_2 \mathcal{A}} \left[ (m + a_2)^2 (\tau_1^2 + \tau_2^2) + (n + a_1)^2  - 2 (n + a_1)(m + a_2) \tau_1 \right]
\eea

After summing over all integers, it is clear that modular transformations transform between the different topological sectors as follows:
\bea
\mathcal{T}: \qquad (a_1,a_2) &\rightarrow& (a_1 + a_2,a_2) \nn
\mathcal{S}: \qquad (a_1,a_2) &\rightarrow& (a_2,a_1)
\label{modular}
\eea
where $a_1$ and $a_2$ are defined modulo an integer. Note that if we include any of the antiperiodic sectors, modular invariance forces us to include the other two.

We note that the above relations will also cause extra degeneracies to arise for special values of $\tau$. For example, we consider the square torus $\tau = i$ in the main text, which satisfies $\tau = -1/\tau$. Since the full spectrum must be invariant under $\mathcal{S}$, the $(1/2,0)$ and $(0,1/2)$ sectors are degenerate. We also consider the triangular torus $\tau = e^{i \pi/3}$, which satisfies $\tau = -1/(\tau - 1)$. Then the invariance of the full spectrum under $\mathcal{T}^{-1}\mathcal{S}$ means all three nontrivial sectors have exactly degenerate spectra.

In the main text, we gave universal results for the cylindrical limit, which is $\tau_1 = 0$, $\tau_2 \rightarrow \infty$. This can be done either by considering formulating the problem on the cylinder to begin with, or by taking the limit of the special function in Eq.~(\ref{specfunc}). The limit requires needs to be taken carefully because of the competing dependencies $g_{s}^{(d)}$ on $\tau_2$, but by similar manipulations to the above derivation the limits can be extracted. For the two cases we use in the text, the limits are given by
\bea
g_{1/2}^{(2)}(\Delta,\tau = i\infty) = \int_{1}^{\infty} d \lambda \lambda^{-1/2} \left( \exp\left( - \frac{L^2 \Delta^2}{4 \pi \lambda} \right) \vartheta_3(\pi a_1,\lambda) - 1 \right) \nn
+ \int_{1}^{\infty} d \lambda \lambda^{-1} \exp\left( - \frac{\lambda L^2 \Delta^2}{4 \pi} - \pi \lambda a_1^2 \right) \vartheta_3(- i \pi\lambda a_1,\lambda) - 2 \label{cylgap}
\eea
and
\bea
g_{-1/2}^{(2)}(\Delta,\tau = i\infty) = - \frac{\tau_2^2}{2 \pi} \Bigg\{ \int_{1}^{\infty} d \lambda \lambda^{1/2} \left( \exp\left( - \frac{L^2 \Delta^2}{4 \pi \lambda} \right) \vartheta_3(\pi a_1,\lambda) - 1 + \frac{L^2 \Delta^2}{4 \pi \lambda} \right) \nn
+ \int_{1}^{\infty} d \lambda \lambda^{-5/2} \exp\left( - \frac{\lambda L^2 \Delta^2}{4 \pi} - \pi \lambda a_1^2 \right) \vartheta_3(- i \pi\lambda a_1,\lambda) - \frac{2}{3} + \frac{L^2 \Delta^2}{2\pi} \Bigg\}. \label{cylground}
\eea
Here we have used the Jacobi theta function, defined as
\beq
\vartheta_3(i \pi a,b) \equiv \sum_{n \in \mathbb{Z}} \exp \left( - \pi b n^2  - 2 \pi a n \right). 
\eeq
The gap on the cylinder can be obtained by using Eq.~(\ref{cylgap}) in the gap equation (\ref{gapeqn}). From Eq.~(\ref{cylground}), we see that $g_{-1/2}^{(2)}$ diverges in the cylindrical limit. This is related to the area dependence of the ground state energy, which is given by
\beq
E_0 = \frac{2 \pi N}{\tau_2 L} g_{-1/2}^{(2)}(\Delta,\tau) + \frac{N (s - s_c ) }{2} \tau_2 L^2 \Delta^2. 
\eeq 
The ground state energy is proportional to $\tau_2$ in this limit, diverging as the area becomes infinite. However, the dimensionless ground state energy density, $L E_0/\tau_2 = L^3 E_0 / \mathcal{A}$, remains a universal function of $s - s_c$, which we compute in the main text.

\section{$1/N$ corrections}
\label{1/n}
Here we mention the form of the leading $1/N$ corrections, following a similar notation to Ref.~\onlinecite{DPSS12}. First we need to calculate the critical coupling $s_c$ to order $1/N$. This is done by solving the infinite-volume gap equation (\ref{gap1}), where we write the infinite volume saddle point as $i \tilde{\lambda} = r + i \lambda/\sqrt{N}$:
\beq
\frac{r}{u} = \frac{s_c}{2} + \int \frac{d^{d+1} p}{\left(2 \pi\right)^{d+1}} \frac{1}{p^2 + r}. \label{infgap}
\eeq
The coupling $s_c$ should be tuned so that the the energy gap in an infinite volume vanishes. We do this by working backwards: we first calculate the energy gap as a function of $r$, then tune $r$ such that the energy gap vanishes, and finally define $s_c$ through Eq. (\ref{infgap}). From the action (\ref{S3}), the relevant self-energy diagram corrections to the $z_{\alpha}$ propagator are
\bea
&& G^{-1}_{\infty}(p) = p^2 + r + \frac{1}{N}\int \frac{d^{d+1} q}{\left(2 \pi\right)^{d+1}}
 \frac{1}{\Pi_{\infty}(q,r)}\frac{1}{((p+q)^2 + r)} \nn
&&- \frac{1}{N}\frac{1}{\Pi_{\infty}(0,r)} \int \frac{d^{d+1} q_1}{\left(2 \pi\right)^{d+1}}\frac{d^{d+1} q_2}{\left(2 \pi\right)^{d+1}}
 \frac{1}{\Pi_{\infty}(q_1,r)}\frac{1}{(q^2_2 + r)^2 ((q_1 + q_2)^2 + r)}, \label{GinfN}
\eea
where we have the inverse $\lambda$ propagators in an infinite volume:
\bea
\Pi_{\infty}(q,r) &=& \int \frac{d^{d+1} q}{\left(2 \pi\right)^{d+1}} \frac{1}{(q^2 + r)((p+q)^2 + r)}.
\label{piinf}
\eea
The critical point is given by $G^{-1}_{\infty}(0) = 0$, so to order $1/N$,
\bea
r &=& -\frac{1}{N}\int \frac{d^{d+1} q}{\left(2 \pi\right)^{d+1}}
 \frac{1}{\Pi_{\infty}(q,0)}\frac{1}{(p+q)^2} + \frac{1}{N}\frac{1}{\Pi_{\infty}(0,0)} \int \frac{d^{d+1} q_1}{\left(2 \pi\right)^{d+1}}\frac{d^{d+1} q_2}{\left(2 \pi\right)^{d+1}} \frac{1}{\Pi_{\infty}(q_1,0)}\frac{1}{q_2^4 (q_1 + q_2)^2} \nn
 &=& \frac{1}{N} \frac{1}{\Pi_{\infty}(0,0)} \int \frac{d^{d+1} q_1}{\left(2 \pi\right)^{d+1}}\frac{1}{\Pi_{\infty}(q_1,0)} \int\frac{d^{d+1} q_2}{\left(2 \pi\right)^{d+1}} \frac{1}{q_2^4}\left(\frac{1}{(q_1 + q_2)^2} - \frac{1}{q_1^2} \right).
\eea
Note that $\Pi_{\infty}(0,0)$ is really infrared divergent, but it can be regulated, and it cancels out of physical values \cite{DPSS12}. In this case, using dimensional regularization, we notice that
\beq
\int\frac{d^{d+1} q_2}{\left(2 \pi\right)^{d+1}} \frac{1}{q_2^4}\left(\frac{1}{(q_1 + q_2)^2} - \frac{1}{q_1^2} \right) = \frac{\left( q_1^2 \right)^{(d+1)/2-3}}{(4 \pi)^{(d+1)/2}} \frac{\Gamma(\frac{d-3}{2})\Gamma(\frac{d-1}{2})}{\Gamma(d-2)} \xRightarrow{d = 2} 0.
\eeq
So $r$ is of order $1/N^2$ at the critical point in two spatial dimensions, and from Eq. (\ref{infgap}), the critical coupling is of order $1/N^2$ in dimensional regularization. Therefore, there is no $1/N$ correction to the finite volume gap equation (\ref{gap1}).

We can now calculate the self-energy corrections to the $z_{\alpha}$ in a finite volume. These are given by a similar calculation to the one above, but now with loop sums,
\bea
&& G^{-1} (k, i\omega) = \omega^2 + k^2 + \Delta^2 + \frac{1}{N \mathcal{A}} \sum_{q} \int \frac{d \Omega}{2 \pi}
 \frac{D (q, i \Omega)}{((\omega+ \Omega)^2 + (k+q)^2 + \Delta^2)} \nn
&&- \frac{D(0,0)}{N \mathcal{A}^2} \int \frac{d \Omega_1 d \Omega_2}{4 \pi^2} \sum_{q_1, q_2}
\frac{D (q_1, i\Omega_1)}{(\Omega_2^2 + q_2^2 + \Delta^2)^2 ((\Omega_1 + \Omega_2)^2 + (q_1+q_2)^2 + \Delta^2)}.\label{G1N}
\eea
The spectrum is then obtained by solving $G^{-1} (k, E(k)) = 0$.

There are also $1/N$ corrections to the singlet states. To compute these, we need the nonlinear terms in the effective action for $\lambda$, (\ref{efflam}). To order $1/N$ these are
\bea
\mathcal{S}_1 = -\frac{i}{6\sqrt{N}}\frac{1}{\mathcal{A}^3}\sum_{k_1,k_2,k_3} \int \prod_{i=1}^3 \left(\frac{d \omega_i}{2 \pi}\right) K_3(p_1,p_2,p_3) \lambda(p_1)\lambda(p_2),\lambda(p_3) \delta(p_1 + p_2 + p_3) \nn
- \frac{1}{24N} \frac{1}{\mathcal{A}^4}\sum_{k_1,k_2,k_3,k_4} \int \prod_{i=1}^4 \left(\frac{d \omega_i}{2 \pi}\right) K_4(p_1,p_2,p_3,p_4) \lambda(p_1)\lambda(p_2)\lambda(p_3)\lambda(p_4) \delta(p_1 + p_2 + p_3 + p_4)
\eea
using condensed notation where $p_i$ represents $k_i$ and $\omega_i$. The functions in the action are given by
\bea
K_3 &=& 2\sum_{q} \int d\Omega \frac{1}{(\Omega^2 + |q|^2 + \Delta^2)((\Omega+\omega_1)^2 + |q+k_1|^2 + \Delta^2)((\Omega-\omega_2)^2 + |q-k_2|^2 + \Delta^2)}, \nn
K_4 &=& 6\sum_q \int d\Omega \frac{1}{(\Omega^2 + |q|^2 + \Delta^2)((\Omega+\omega_1)^2 + |q+k_1|^2 + \Delta^2)} \nn
&\times&\frac{1}{((\Omega+\omega_1+\omega_2)^2 + |q+k_1+k_2|^2 + \Delta^2)((\Omega-\omega_4)^2 + |q-k_4|^2 + \Delta^2)}
\eea
The propagator for $\lambda$ can then be computed from these interactions terms. One finds that the order $1/N$ correction to the inverse propagator is given by
\bea
D^{-1}(k,i\omega) &= &\Pi(k,i\omega) + \frac{1}{2N\mathcal{A}}\sum_q \int\frac{d\Omega}{2\pi} \left[ K_3(k,q,|k+q|)\right]^2 D_0(|k+q|,i\omega + i\Omega)D_0(q,i\Omega) \nn
&+& \frac{1}{2N\mathcal{A}} \frac{K_3(k,-k,0)}{\Pi(0,0)} \sum_q \int \frac{d\Omega}{2\pi}K_3(q,-q,0)D_0(q,i\Omega) \nn
&+& \frac{1}{6N\mathcal{A}} \sum_q \int \frac{d \Omega}{2\pi} \left[ K_4(k,q,-k,-q) + 2 K_4(k,-k,q,-q) \right] D_0(q,i\Omega),
\eea
and the spectrum of the singlet states is found by solving $D^{-1}(k,E(k)) = 0$.

\bibliography{on}
\end{document}